\documentclass[aps,prb,twocolumn,superscriptaddress]{revtex4-1}
\usepackage{amsfonts}
\usepackage{graphicx}
\usepackage{epstopdf}
\usepackage{epsfig}
\usepackage{amsmath}
\usepackage{bm}
\usepackage{color}
\usepackage{makecell}
\usepackage{array}
\usepackage{booktabs}
\usepackage{tabularx}

\begin{document}

\title{Phase locking of a pair of nano ferromagnetic oscillators on a topological insulator}

\author{Cheng-Zhen Wang}
\affiliation{School of Electrical, Computer and Energy Engineering, Arizona State University, Tempe, Arizona 85287, USA}

\author{Hong-Ya Xu}
\affiliation{School of Electrical, Computer and Energy Engineering, Arizona State University, Tempe, Arizona 85287, USA}

\author{Nicholas D. Rizzo}
\affiliation{Northrop Grumman Corporation, Linthicum, Maryland 21090, USA}

\author{Richard A. Kiehl}
\affiliation{School of Electrical, Computer and Energy Engineering, Arizona State University, Tempe, Arizona 85287, USA}

\author{Ying-Cheng Lai} \email{Ying-Cheng.Lai@asu.edu}
\affiliation{School of Electrical, Computer and Energy Engineering, Arizona State University, Tempe, Arizona 85287, USA}
\affiliation{Department of Physics, Arizona State University, Tempe, Arizona 85287, USA}

\date{\today}

\begin{abstract}

We investigate the magnetization dynamics of a pair of ferromagnetic insulators
(FMIs) deposited on the surface of a topological insulator (TI). Due to the 
nonlinear nature of the underlying physics and intrinsic dynamics, the FMIs 
can exhibit oscillatory behaviors even under a constant applied voltage. The 
motion of the surface electrons of the TI, which obeys relativistic quantum 
mechanics, provides a mechanism of direct coupling between the FMIs. In 
particular, the spin polarized current of the TI surface electrons can affect 
the magnetization of the two FMIs, which in turn modulates the electron 
transport, giving rise to a hybrid relativistic quantum/classical nonlinear 
dynamical system. We find robust phase and anti-phase locking between the 
magnetization dynamics. As driving the surface electrons of a TI only requires 
extremely low power, our finding suggests that nano FMIs coupled by a spin 
polarized current on the surface of TI have the potential to serve as the 
fundamental building blocks of unconventional, low-power computing paradigms.

\end{abstract}
\maketitle

\section{Introduction} \label{sec:intro}

There has been an increasing need to develop unconventional computing
paradigms to deal with special tasks, such as rapid image recognition, with
which conventional digital computing based on integrated circuits finds
fundamental difficulties. Networks of nanoscale oscillators could provide
the needed paradigm for such tasks, where extremely fast image recognition
could potentially be realized with non-Boolean networks in which processing
is done by local operations using analog techniques naturally suited to the
task. Due to the inevitable power dissipation of the oscillators, it is desired
to develop ultra-small oscillators based on a highly energy-efficient physical
mechanism to realize energy efficient computing.

The history of computing with oscillators dates back to Goto~\cite{Goto:1954}
and von Neumann~\cite{Neumann:1957} who proposed to represent Boolean logic
states by the electrical phase of an oscillator, rather than by the voltage
or current level. Almost forty years later, an energy efficient
implementation of this scheme based on single-electron tunneling oscillators,
referred to as tunneling phase logic, was proposed~\cite{KO:1995,OK:1996}.
The synchronization behavior of such pulse coupled oscillators opens the
possibilities for non-Boolean computing~\cite{Izhikevich:1996} as well. Coupled
nano-oscillators are also promising for combinatorial optimization problems
such as vertex coloring of graphs~\cite{WJLC:2011} with applications in
scheduling~\cite{Leighton:1979}, resource allocation~\cite{WSN:1991} and
other computationally difficult (NP-hard) problems~\cite{PSJDR:2017}. While
tunneling phase logic was shown to be capable of both Boolean
functions~\cite{LAK:1999} and non-Boolean image processing
operations~\cite{YKC:2001}, its physical mechanism was found to be too
sensitive to thermal noise~\cite{LK:2003} for practical applications, calling
for schemes to suppress noise~\cite{nezlobin2003phase} and alternative 
mechanisms for developing more robust 
nano-oscillators~\cite{kiehl2016information}.

In general, the ability to control and manipulate magnetization dynamics is 
essential to developing spintronic memory, logic, and sensing nanodevices. A 
mechanism that has been extensively studied theoretically and experimentally is
spin-transfer torque~\cite{slonczewski1996current,berger1996emission}, which is
based on the transfer of the spin angular momentum between a spin current flow
and the local magnetization of a ferromagnetic layer. The mechanism can be
exploited to develop, e.g., switching and steady precession of spin torque
oscillators (STOs)~\cite{ralph2008spin,brataas2012current}. The dynamics
of precession of a single STO provide the basis for synchronizing a number of
STOs~\cite{kiselev2003microwave,rippard2004direct,krivorotov2005time}, which
has applications such as microwave power generation and sensing. Phase locking 
of two STOs has been achieved experimentally in spin torque devices with 
multiple nanocontacts, in which the magnetization in all the nanocontact 
regions can be locked at the same phase via a propagating spin 
wave~\cite{kaka2005mutual,mancoff2005phase,slavin2006theory}. Phase
locking of STOs through coupled electrical circuits has also been studied in
an array of STO nanopillars that can be electrically connected in series or in
parallel~\cite{GCF:2006,PZA:2007,GGCF:2008,LZZH:2010,Sanietal:2013,TBPDIL:2017}.
In this case, the AC current produced by each individual oscillator leads to
a feedback among the STOs, thereby realizing synchronization. In addition,
synchronization can be achieved through magnetic dipolar coupling in 
perpendicular-to-plane polarized STOs~\cite{chen2016phase}. Local 
synchronization between vortex-based STOs interacting with each other can occur through the mediation of closely spaced antivortices~\cite{Ruotoloetal:2009}. 
Recently, spin-torque and spin-Hall nano-oscillators~\cite{Chenetal:2016}
have gained attention for their potential applications in various non-Boolean
computing~\cite{SAR:2012} including image processing~\cite{HCCPR:2012},
associative memory, pattern recognition~\cite{CP:2013,Nikonovetal:2015,
FMYSR:2015}, and spatiotemporal wave computing~\cite{MHK:2014}. Especially, 
the spin Hall effect~\cite{Demidovetal:2012,LPRB:2012,LLU:2013,GCLGAF:2014} 
was exploited to experimentally realize synchronization of STOs driven 
by a pure spin current through microwave driving~\cite{Demidovetal:2014} and 
a method to synchronize multiple STOs without requiring any external AC 
excitation was proposed~\cite{EBY:2015}. Existing studies on STOs have been 
focused primarily on nanocontact spin valves and magnetic tunnel junction 
pillar structures. While the junction structures appear more promising for 
microwave power generation because of their high junction resistance and a 
larger magnetoresistance, nanocontact spin valves are more promising for mutual
phase-locking among multiple STOs because of their better interdevice coupling 
geometry~\cite{kaka2005mutual,mancoff2005phase,slavin2006theory,QV:2015}.

In this paper, motivated by the growing interest in exploiting topological
quantum materials for achieving novel charge transport and efficient electrical
control of magnetization in spintronics applications, we investigate
whether it is possible to realize phase locking of nanoscale magnetic
oscillators coupled via some topological mechanism, e.g., through a
topologically protected current. This has the potential to lead to
highly efficient, low power nano-oscillators as the fundamental building
blocks of unconventional computing paradigms. To be concrete, we consider
the prototypical setting of a pair of ferromagnetic insulators (FMIs) on the
surface of a three-dimensional (3D) topological insulator (TI). A 3D TI
possesses an insulating bulk but hosts chiral metallic channels on its
surface, where electrons are described as massless Dirac fermions with
spin-momentum locking~\cite{HK:2010,QZ:2011}, resulting in large spin-charge
conversion efficiency~\cite{Shiomietal:2014,Deoranietal:2014,Jamalietal:2015,
Rogasetal:2016}. The locking provides an effective mechanism to control 
FMI magnetization~\cite{Fanetal:2016,WDBKBOY:2015}, and a large figure of
merit for charge-spin conversion has been experimentally
realized~\cite{Mellniketal:2014,Fanetal:2014}. For a single FMI
deposited on the top of a 3D TI, the exchange coupling between the
magnetization and the surface state of TI can lead to nonlinear magnetization
evolution but the spin-momentum locking of the surface current of the TI is
preserved, and this can lead to phenomena such as anomalous magnetoresistance,
unconventional transport behaviors~\cite{YTN:2010,WPC:2010}, and
magnetization switching due to Hall current induced effective anisotropy
field~\cite{GF:2010,YZN:2010,Yokoyama:2011}. Quite recently, steady
self oscillations in the FMI/TI heterostructure were
uncovered~\cite{SDK:2014,DLSK:2015} and explained~\cite{NAFVKM:2017}, and
a number of nonlinear dynamical behaviors were
studied~\cite{WXL:2016,WXL:2018}. Here, we apply a DC voltage to the TI and
place the two FMIs on the top of the TI in series. We first consider the
case where the distance between the two FMIs is larger than the de Broglie
wavelength so that quantum interference between the two FMIs can be
neglected. As a result, the surface electronic states provide the only
mechanism that couples the two FMIs. We calculate the average spin
of the electron flow in each heterostructure interface by solving the
quantum transmission. The effective spin field, when combined with the
magnetic anisotropy of the FMIs, can lead to self oscillations of
the magnetization vectors of the FMIs, {\em even when the external driving
is DC}. The oscillations in turn can modulate the electron transmission
periodically, effectively making the current time varying. The resulting AC
current provides the needed coupling between the two FMIs for phase locking.
We then study the case where there is quantum interference between the two
FMIs and find robust phase locking. The topologically coupled FMI system
thus represents a class of highly efficient, low power nanoscale coupled oscillators,
which can potentially serve as the fundamental building blocks for
unconventional computing paradigms.

\section{Model and solution method} \label{sec:model}

\begin{figure}
\centering
\includegraphics[width=\linewidth]{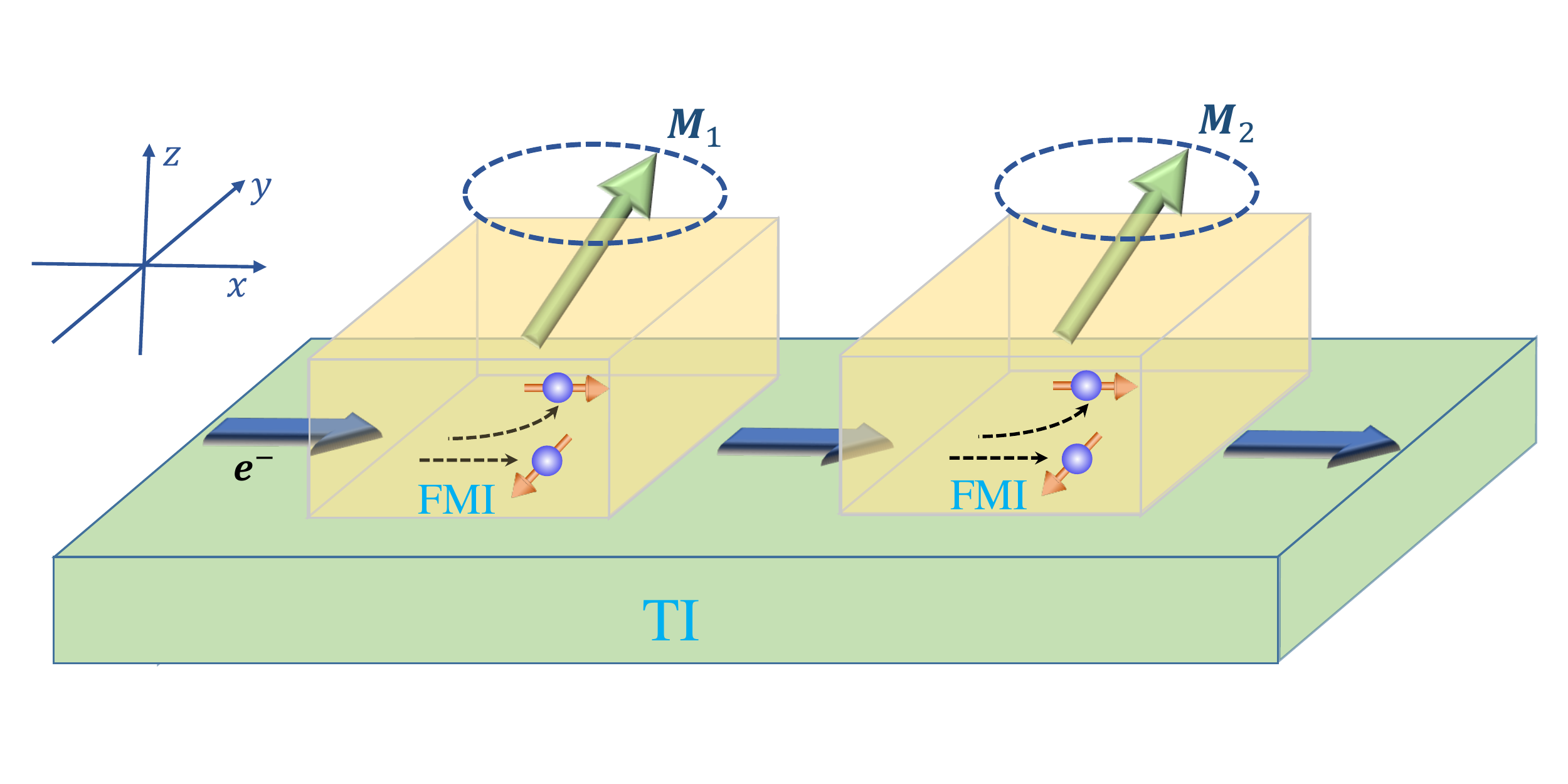}
\caption{ {\bf Schematic illustration of two FMIs on the top of a TI}.
The two yellow blocks represent the FMIs and the green block underneath is 
the TI. The green arrows in the yellow blocks are the magnetization vectors 
$\boldsymbol{M}$ and the blue arrows on the surface of green blocks represent 
certain electron incident direction in the magnetic free region. The small 
red arrows with the electrons (blue spheres) denote the spin direction and 
the black arrows indicate the direction of electron motion.}
\label{fig:Schematic_setup}
\end{figure}

Figure~\ref{fig:Schematic_setup} shows schematically the system configuration
of two FMIs deposited on the top of a TI~\cite{RKS:2015,Hellmanetal:2017},
which can be realized using material combinations such as
Bi$_2$Se$_3$/YIG (Y$_3$Fe$_5$O$_{12}$)~\cite{lang2014proximity},
Bi$_2$Te$_3$/GdN~\cite{kandala2013growth},
Bi$_2$Se$_3$/EuS~\cite{wei2013exchange,yang2013emerging},
and Bi$_2$Se$_3$/Cr$_2$Ge$_2$Te$_4$~\cite{alegria2014large}.
(Appendix~\ref{App_A_exp} provides more details about possible experimental 
realization.) The dynamical variable of each FMI is its macroscopic 
magnetization vector $\boldsymbol{M}$. For the TI, a topologically protected, 
spin polarized current flows through the surface, where the spin
is perpendicular to the current flow direction. The spin and magnetization
are coupled via proximity interaction. The magnetization can affect
the spin distribution and hence the electron transport behavior.
Simultaneously, the average spin will act as an effective magnetic field
to influence the dynamics of the FMIs. Even with constant voltage driving,
the magnetization vectors of the FMIs can exhibit oscillations. Intuitively,
because the spin polarized current is common to both FMIs, it serves as a 
kind of coupling between the two FMIs. Specifically, the magnetization
of the first FMI can affect the current, which in turn alters the effective
magnetic field acting on the second FMI, impacting its dynamics, and vice
versa. As a result, phase locking can arise.

To develop a computational model, we assume that the magnetization precession
period is much longer than the time it takes for electronic transport through 
the FMI/TI interface. For simplicity, we first ignore the quantum interference 
effect or any other indirect interaction between the two FMIs. (The effect
of quantum interference will be discussed in Sec.~\ref{subsubsec:WQI}.)  
We solve the time-independent Dirac equation for the electrons at the two 
interfaces separately, taking into account the proximity effect. In 
particular, the low-energy electronic behavior of the TI surface states is 
described by the effective Dirac Hamiltonian~\cite{SDK:2014}
\begin{equation} \label{eq:Hamilton}
H = \hbar v_F (\boldsymbol{\sigma} \times \boldsymbol{k}) \cdot
\hat{\boldsymbol{z}} - \xi \boldsymbol{M} \cdot \boldsymbol{\sigma} - U,
\end{equation}
where $\boldsymbol{p}=\hbar\boldsymbol{k}=-i\hbar(\partial_x,\partial_y,0)$
is the two-dimensional (2D) electron momentum operator,
$\boldsymbol{\sigma} = (\sigma_x, \sigma_y, \sigma_z)$ are the Pauli matrices
for electron spin, $\hat{\boldsymbol{z}} = (0, 0, 1)$ is the unit vector
normal to the TI surface, and $v_F$ is the electron Fermi velocity, as shown
in Fig.~\ref{fig:Schematic_setup}. The second term in Eq.~(\ref{eq:Hamilton})
represents the energy of exchange interaction between an electron and the
proximate FMI, with $\xi$ being the coupling coefficient. The last term is
the external bias applied to the interface regions. 
When quantum interference is neglected, we can treat the two FMI/TI 
heterostructures separately. Especially, we solve Eq.~(\ref{eq:Hamilton}) 
by taking $\boldsymbol{M} = \boldsymbol{M}_1$ in the first heterostructure 
region and match the wavefunctions without the need to consider the influence 
of the second heterostructure. Similarly, the second heterostructure 
can be treated without considering any influence from the first one. The 
formulas for calculating the electron transmission through one FMI/TI 
heterostructure can be found in Appendix~\ref{App_C_Iteration}. When quantum 
interference is taken into account, we solve Eq.~(\ref{eq:Hamilton}) for the 
two FMIs as a whole - see Sec.~\ref{subsubsec:WQI} and Appendix~\ref{App_D_QIF}
for details.

The energy eigenvalues of Eq.~(\ref{eq:Hamilton}) are
\begin{displaymath}
E_{\pm}=\pm\sqrt{(v_Fp_x + \xi M_y)^2 +(v_Fp_y - \xi M_x)^2+\xi^2 M_z^2} - U,
\end{displaymath}
where the ``$\pm$'' signs correspond to the conduction ($+$) and valence ($-$)
bands, and $\boldsymbol{p}=\hbar \boldsymbol{k}$ is the electron momentum. We
see that the in-plane $(x,y)$ magnetization components can lead to a
displacement in the momentum space. Especially, the momentum displacement in
the $y$ direction can lead to a Hall current in that direction. Besides, the
perpendicular component of the magnetization vector can open up a gap between
the Dirac cones, contributing an additional Hall current in $y$. The first
kind of Hall current plays the role of effective anisotropy, while the second
kind is responsible for anti-damping. The two kinds of Hall current can lead
to self oscillations of magnetization~\cite{SDK:2014,DLSK:2015,NAFVKM:2017}.

For each FMI, the conductance through one FMI/TI heterostructure can be
calculated from the Landauer-Buttiker 
formalism~\cite{WPC:2010,SDK:2014,Datta:book}:
\begin{align}\label{Eq:conductance}
G = \frac{Ee^2 L_w}{2\pi^2 \hbar^2 v_F}\int^{\frac{\pi}{2}}_{-\frac{\pi}{2}}
T_{\boldsymbol{M}}(E, \theta) \cos \theta d\theta.
\end{align}
where E is electron Fermi energy, 
$T_{\boldsymbol{M}}(E, \theta) = |t|^2$ is the transparency through one
FMI/TI barrier, $\theta$ is the electron incident angle in the $(x,y)$ plane,
and $-e$ is the electron charge. For two coupled FMIs, their conductances
$G_1$ and $G_2$ determine the voltage partition between them:
\begin{align}\label{eq:voltage_partition}
V_1 = \frac{G_2}{G_1 + G_2}V \ \ \mbox{and} \ \ V_2 = \frac{G_1}{G_1 + G_2}V.
\end{align}
The current density is given by
\begin{align} \label{Current_Density}
J_x &= \frac{V_1 G_1}{L_w}= \frac{Ee^2 V_1}{2\pi^2 \hbar^2 v_F}
\int^{\frac{\pi}{2}}_{-\frac{\pi}{2}} T_{\boldsymbol{M}}(E, \theta)
\cos{\theta}d\theta.
\end{align}
From the current definition~\cite{Yokoyama:2011,SDK:2014} 
$\hat{J}=-e \nabla_{\boldsymbol{p}}H = -ev_F(-\hat{\sigma}_y, 
\hat{\sigma}_x)$, we can get the mean spin polarization density for the 
first FMI as
\begin{align}
\langle\sigma_{y}\rangle_1 &= J_x/ev_F, \\
&= \frac{EeV_1}{2\pi^2 \hbar^2 v^2_F}
\int^{\frac{\pi}{2}}_{-\frac{\pi}{2}} T_{\boldsymbol{M}}(E, \theta)
\cos{\theta}d\theta,
\end{align}
The following equality:
\begin{displaymath}
T_{\boldsymbol{M}}(E,\theta)\cos{\theta}=-\psi^{\dagger}\sigma_y \psi,
\end{displaymath}
where $\psi$ is the electron wavefunction, can be used in our derivation of 
the average spin density (a proof of this equality is presented in 
Appendix~\ref{App_C_Iteration}). Specifically, using the equality, we have   
\begin{align} \label{eq:sigma_y_1}
\langle\sigma_{y}\rangle_1 &=-\frac{EeV_1}{2\pi^2 \hbar^2 v^2_F}
\int^{\frac{\pi}{2}}_{-\frac{\pi}{2}} \psi^{\dagger} \sigma_y \psi d\theta \nonumber\\
&= -\frac{EeV_1}{2\pi^2 \hbar^2 v_F^2 d}\int^{d}_0
\int^{\frac{\pi}{2}}_{-\frac{\pi}{2}} \psi^{\dagger} \sigma_y \psi d\theta dx
\end{align}
There are three spin components for each electron at a specific position 
with certain incident angle: $\psi^{\dagger} \sigma_x \psi$, 
$\psi^{\dagger} \sigma_y \psi$, and $\psi^{\dagger} \sigma_z \psi$. 
Once the $y$ component of the spin density is known, the other 
components can be obtained by replacing $\psi^{\dagger} \sigma_y \psi$ in 
Eq.~(\ref{eq:sigma_y_1}) by $\psi^{\dagger} \sigma_x \psi$ and 
$\psi^{\dagger} \sigma_z \psi$. Note that the factor before the integral 
is related to the electron density. We have
\begin{align}
\langle\sigma_{x}\rangle_1 &= -\frac{EeV_1}{2\pi^2 \hbar^2 v_F^2 d}\int^{d}_0\int^{\frac{\pi}{2}}_{-\frac{\pi}{2}} \psi^{\dagger} \sigma_x \psi d\theta dx,\label{eq:sigma_x_1} \\
\langle\sigma_{z}\rangle_1 &= -\frac{EeV_1}{2\pi^2 \hbar^2 v_F^2 d}\int^{d}_0\int^{\frac{\pi}{2}}_{-\frac{\pi}{2}} \psi^{\dagger} \sigma_z \psi d\theta dx. \label{eq:sigma_z_1}
\end{align}
Alternatively, we can get the electron density first and then obtain the 
spin density expression. A detailed discussion is presented in 
Appendix~\ref{App_C_Iteration}.

The mean spin density for the second FMI can be obtained in a similar way.
The effective magnetic field is then given by
\begin{align}
\boldsymbol{B}_{spin} = -\langle \frac{\partial H}{\partial \boldsymbol{M}} \rangle \frac{A_0}{V_0} = \frac{\xi}{a}\langle \boldsymbol{\sigma} \rangle,
\end{align}
where $\langle\boldsymbol{\sigma}\rangle$ is the mean spin density of the
electron flow.

In addition to the effective magnetic field contribution from the electron
spin, there is another term that stems from the magnetic anisotropy of the
material. We assume that the magnetic layer has $z$ hard axis and $x$ easy 
axis. The corresponding anisotropy parameters are $K_z > K_y >K_x = 0$, and 
the density of the magnetic free energy~\cite{SDK:2014} is given by
\begin{align}
F(\boldsymbol{M}) &=F_{an} + F_{spin} \nonumber \\
&= K_x n_x^2 + K_y n_y^2 + K_z n_z^2 - \boldsymbol{M}\cdot \boldsymbol{B}_{spin}(\boldsymbol{M}),
\end{align}
where $\boldsymbol{n} = (n_x, n_y, n_z) = \boldsymbol{M}/|\boldsymbol{M}|$. 
The effective magnetic field due to material anisotropy can be obtained 
via $\boldsymbol{B}_{an} = -\partial F_{an}/\partial \boldsymbol{M}$.

The LLG equations for magnetization dynamics of the two coupled FMIs are
\begin{align}
\frac{d\boldsymbol{n}_1}{dt} = -\gamma \boldsymbol{n}_1 \times \boldsymbol{B}_{eff}^{(1)}(\boldsymbol{n}_1, \boldsymbol{n}_2)
+ \alpha \boldsymbol{n}_1 \times \frac{d\boldsymbol{n}_1}{dt}, \\
\frac{d\boldsymbol{n}_2}{dt} = -\gamma \boldsymbol{n}_2 \times \boldsymbol{B}_{eff}^{(2)}(\boldsymbol{n}_1, \boldsymbol{n}_2)
+ \alpha \boldsymbol{n}_2 \times \frac{d\boldsymbol{n}_2}{dt},\label{eq:LLGEq}
\end{align}
where the normalized magnetization vectors are
$\boldsymbol{n}_1 = \boldsymbol{M}_1/|\boldsymbol{M}|$ and
$\boldsymbol{n}_2 = \boldsymbol{M}_2/ |\boldsymbol{M}|$, $\gamma$ is the
gyromagnetic ratio, and $\alpha$ is the Gilbert damping constant. The
quantities $\boldsymbol{B}_{eff}^{(1)}$ and $\boldsymbol{B}_{eff}^{(2)}$
are the effective magnetic fields for the first and second FMI, respectively,
with $\boldsymbol{B}^{(1)}_{eff} = \boldsymbol{B}^{(1)}_{spin}
+ \boldsymbol{B}^{(1)}_{an}$ and $\boldsymbol{B}^{(2)}_{eff} =
\boldsymbol{B}^{(2)}_{spin} + \boldsymbol{B}^{(2)}_{an}$. The anisotropy
induced effective magnetic fields $\boldsymbol{B}^{(1)}_{an}$ and
$\boldsymbol{B}^{(2)}_{an}$ are different for the two FMIs, leading to
different oscillation frequencies for the two FMIs under the same applied
voltage. The spin induced effective magnetic field of the first
heterostructure is $\boldsymbol{B}^{(1)}_{spin} =
\xi\langle \boldsymbol{\sigma}\rangle_1/a \sim V_1$, where $V_1$
is determined by the conductances of both heterostructures via voltage
partition with the same longitudinal current in the $x$ direction. The
conductances are related to the magnetization vectors $\boldsymbol{M}_1$
and $\boldsymbol{M}_2$. Similarly, the spin induced effective magnetic field
in the second heterostructure is $\boldsymbol{B}^{(2)}_{spin} \sim V_2$,
which is related to $\boldsymbol{M}_1$ and $\boldsymbol{M}_2$ in the same
way as for the first heterostructure. The magnetization vectors of the two
FMIs are thus effectively coupled together via the common current on the
surface of the TI.

The coupled magnetization dynamics can be solved by an iterative procedure.
Firstly, with the magnetization vectors $\boldsymbol{M}_1$ and
$\boldsymbol{M}_2$ of the two FMIs, we solve the Hamiltonian 
(\ref{eq:Hamilton}) to obtain the corresponding electron 
wavefunctions in the two heterostructures. Secondly, from the wavefunctions, 
we calculate the conductance [Eq.~(\ref{Eq:conductance})] and the average spin. 
Using the common coupling current in the $x$ direction through the two 
heterostructures, we carry out a simple voltage partition 
[Eq.~(\ref{eq:voltage_partition})]. Thirdly, we calculate the spin density
[Eqs.~(\ref{eq:sigma_y_1}-\ref{eq:sigma_z_1})] and obtain the effective 
magnetic field by spin, which affects the magnetization dynamics. These steps 
are repeated to obtain the time evolution of the magnetization vectors.
A flow chart illustrating the iterative method for the two cases where 
quantum interference is absent and present, respectively, is provided in 
Appendix B.

Our simulation parameters are the following. Each magnet is assumed to have 
the dimension of $d$ (length)$\times L_w$ (width) $\times a$ (thickness) 
$=$ $40 \times 90 \times 2.2$ nm$^3$, with hard-axis anisotropy coefficients 
$K_y = 2.0\times 10^5$ erg/cm$^3$ and $K_z = 2.5\times 10^5$ erg/cm$^3$ along 
the $y$ and $z$ axis, respectively. The initial magnetization is $M_0 = 1200$ 
Oe. The Gilbert damping factor is $\alpha = 0.01$. For the TI layer, the Fermi 
velocity of the electron is $v_F = 4.6 \times10^7$ cm/s. The exchange energy 
term is $\xi M_0 = 40$ meV.

\section{Results} \label{sec:results}

\begin{figure}
\centering
\includegraphics[width=\linewidth]{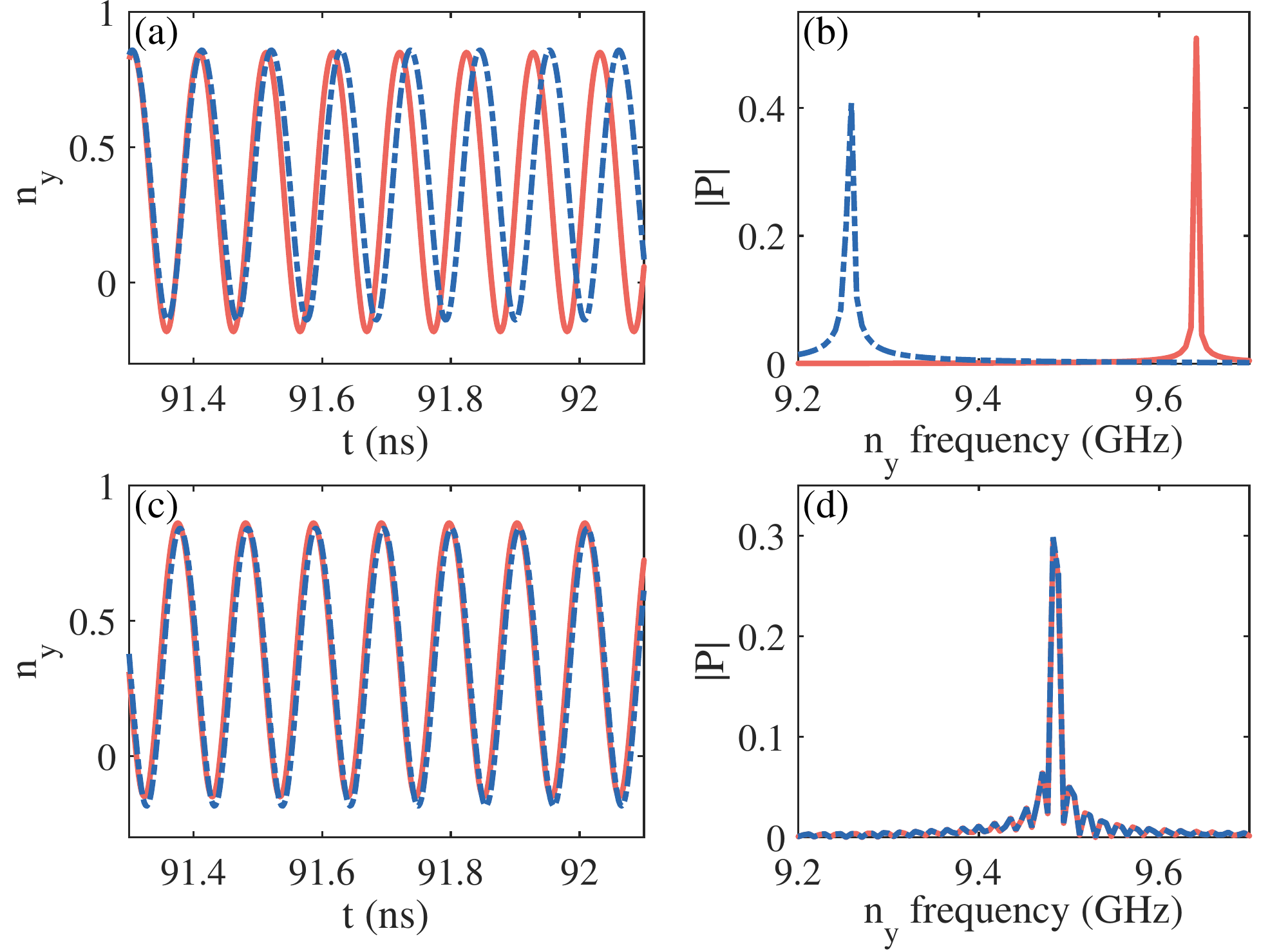}
\caption{ {\bf Phase locking between magnetization vectors of two FMIs}.
For $K_y = 2 \times 10^{4}$ erg/cm$^3$ and electron energy $100$ meV,
(a) time evolution of the $y$ components of magnetization vectors of the
two isolated FMIs under the same applied voltage of $40$ mV. The red solid 
and blue dashed curves denote the $y$ components of the magnetization of the 
first and second FMI, respectively. (b) Fourier spectra of the time series 
in (a). (c,d) The corresponding results with coupling through the surface 
current of the TI under an applied voltage of $80$ mV. There is phase locking.}
\label{fig:series1}
\end{figure}

\begin{figure}
\centering
\includegraphics[width=\linewidth]{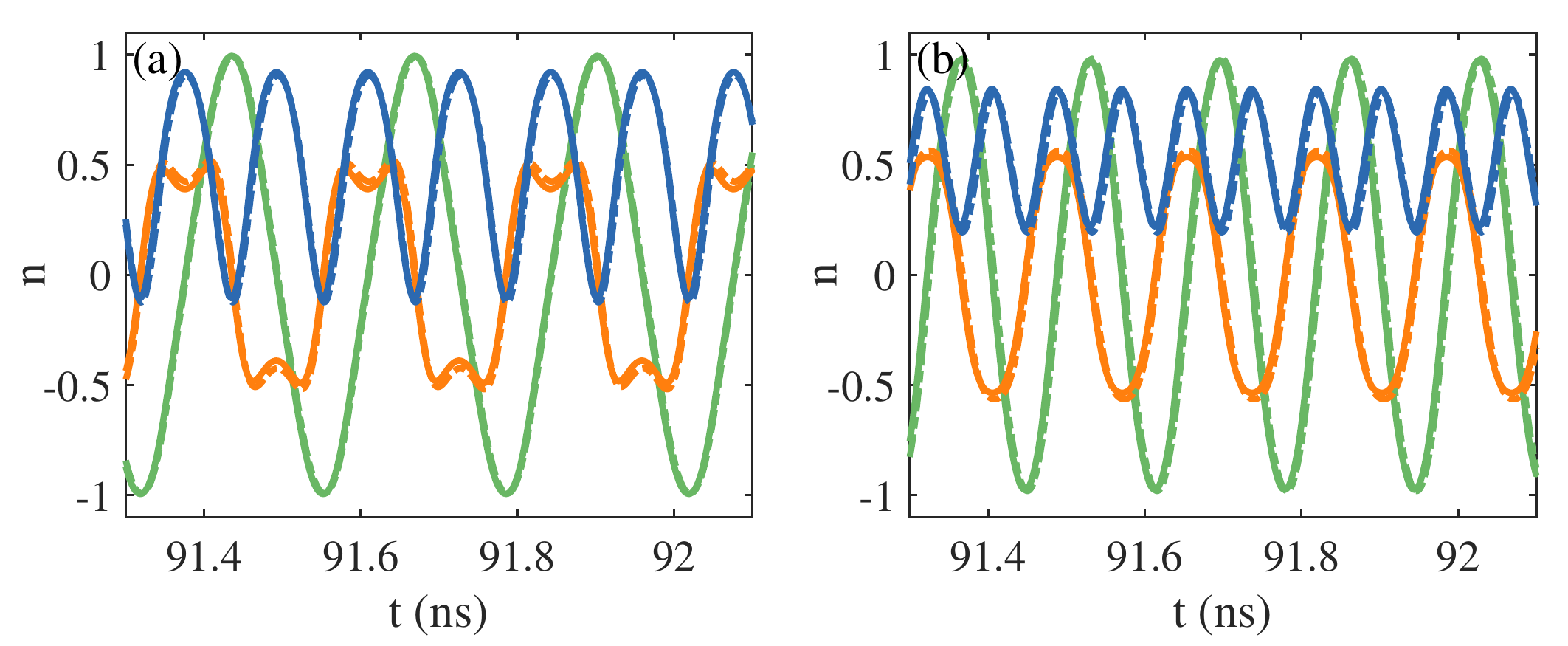}
\caption{ {\bf Robustness of phase locking for different parameter settings}.
For $K_y = 1.5 \times 10^{4}$ erg/cm$^3$ (the anisotropy coefficient in
$y$) and electron energy $100$ meV, (a) phase locking between the two FMIs 
under voltage $100$ mV. The red, blue and green curves denote the $n_x$, 
$n_y$ and $n_z$ components, respectively. The solid and dashed curves are 
for the first and second FMIs, respectively. (b) For applied voltage 
$140$ mV, phase locking behavior when the damping factor is increased 
to $0.02$ (from the value of $0.01$ in Fig.~\ref{fig:series1}).} 
\label{fig:changepara}
\end{figure}

\subsubsection{Phase locking in the absence of quantum interference}
\label{subsubsec:WTQI}

To uncover phase locking for a pair of coupled FMIs in a general setting, we 
assume that the FMIs have different values of anisotropy: one with the values 
listed at the end of Sec.~\ref{sec:model} and the other having an additional 
amount of anisotropy in the $x$ direction with the value of the anisotropy 
coefficient being $K_x = 0.0955 \times 10^5$ erg/cm$^3$. Nonidentical values 
of the anisotropy lead to different oscillation frequencies for the two FMIs 
under the same applied voltage. 

We first consider the case of isolated FMIs by applying the same DC voltage 
on the two FMIs separately. Figure~\ref{fig:series1}(a) shows that the 
magnetization vectors of the isolated FMIs exhibit oscillations at difference 
frequencies, where the solid red and dotted blue curves correspond to the first 
and second FMI, respectively. Note that the two magnetization components 
deviate within $1$ ns (containing several oscillation periods), signifying a
difference in their frequencies due to the difference in the anisotropy.
The frequency difference can also be seen from the Fourier spectra, as
shown in Fig.~\ref{fig:series1}(b). For the second FMI with an additional
value of anisotropy along the $x$ axis, the frequency is lower than that
of the first one. We next introduce coupling by placing the two FMIs in series 
on a TI and letting the current go through the two FMIs, as illustrated in 
Fig.~\ref{fig:Schematic_setup}. The separation between the two FMIs is 
sufficiently large, so that any quantum interference between the two FMIs 
can be neglected. We apply the voltage of $80$ mV. The magnetization 
oscillations will make the current oscillate in time through the proximity 
effect, i.e., modulation of the transmission of electrons. The current induces 
an interaction between the two FMIs through an effective magnetic field due 
to electron spin, leading to phase locking, as shown in 
Fig.~\ref{fig:series1}(c), where the $y$ components of the magnetization 
vectors of the two FMIs evolve with time at the same pace. Phase locking can 
be further demonstrated by the Fourier spectra, as shown in 
Fig.~\ref{fig:series1}(d), where the two oscillatory time series have 
essentially the same peak frequency. We have examined a large number of 
combinations of the parameters such as the amount of anisotropy and damping 
factor, and found robust phase locking in all cases, as exemplified in 
Fig.~\ref{fig:changepara}.

\begin{figure}
\centering
\includegraphics[width=\linewidth]{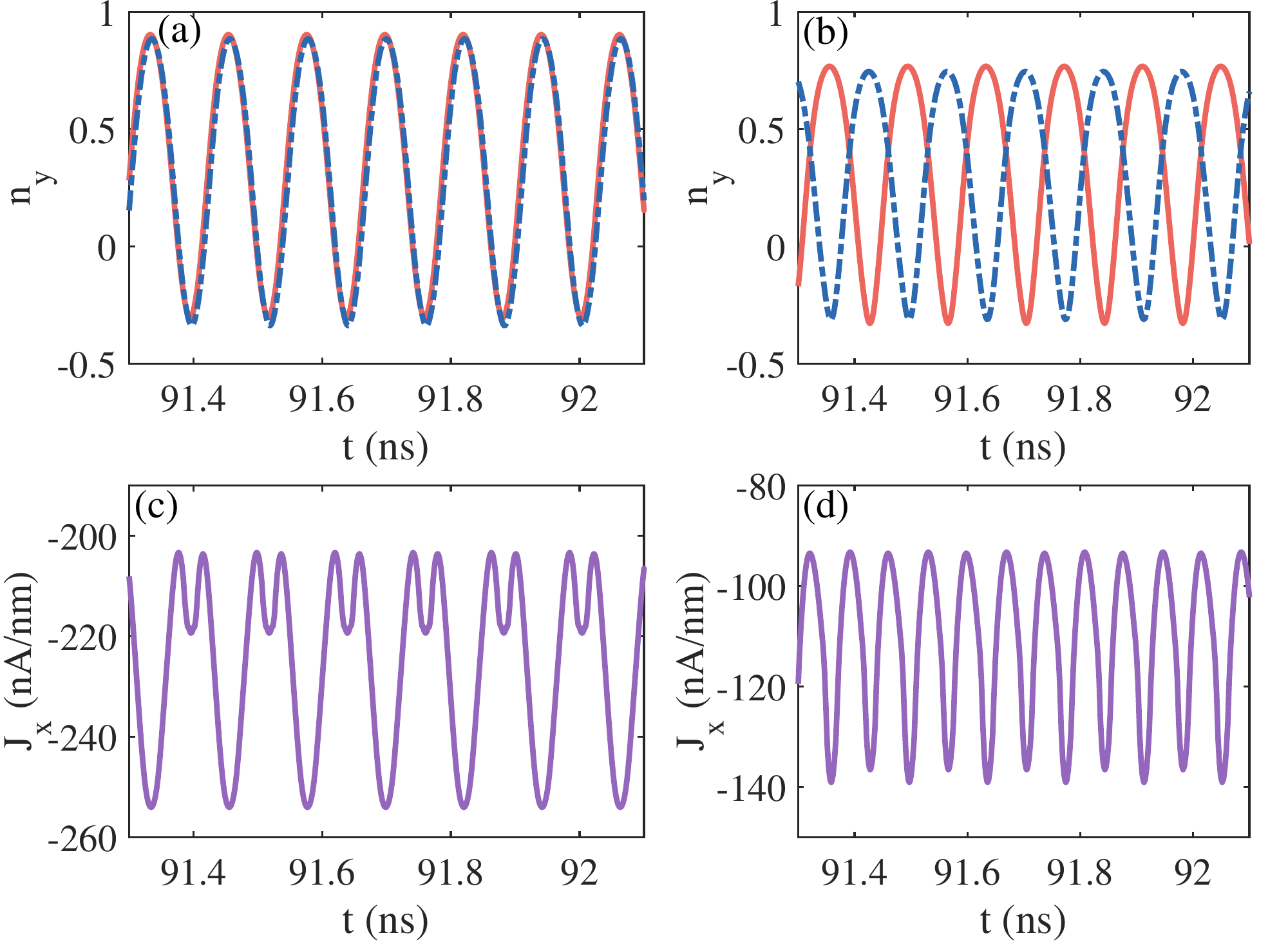}
\caption{ {\bf Phase and anti-phase locking between a pair of
coupled FMIs}. (a) Phase locking between the $y$ components of the
magnetization vectors for $V_0 = 110$ mV and $E = 60$ meV. (b) Anti-phase
locking for $V_0 = 190$ mV and $E = 30$ meV. (c,d) The corresponding
evolutions of the surface current of the TI for cases (a,b), respectively.
Other parameters are the same as those in Fig.~\ref{fig:series1}(c).}
\label{fig:series2}
\end{figure}

We also find persistent phase locking in wide ranges of the applied voltage
and electron Fermi energy. For example, Figs.~\ref{fig:series2}(a,b)
demonstrate phase locking for two cases where the applied voltage
and electron energy are ($110$ mV, $60$ meV) and ($190$ mV, $30$ meV),
respectively, with other parameters being the same as those in
Fig.~\ref{fig:series1}(c). Note that in Fig.~\ref{fig:series2}(a), the
magnetization vectors of the two FMIs are in phase, while there is anti-phase
locking between them in Fig.~\ref{fig:series2}(b). The corresponding surface
current oscillations in the TI are shown in Figs.~\ref{fig:series2}(c,d). In
each case, the primary frequency of the current oscillations is the same as
that of the magnetization oscillations. To our knowledge, the demonstrated
phase and anti-phase locking behaviors enabled by the proximity induced torques
in the FMI/TI systems have not been reported before.

\begin{figure}
\centering
\includegraphics[width=\linewidth]{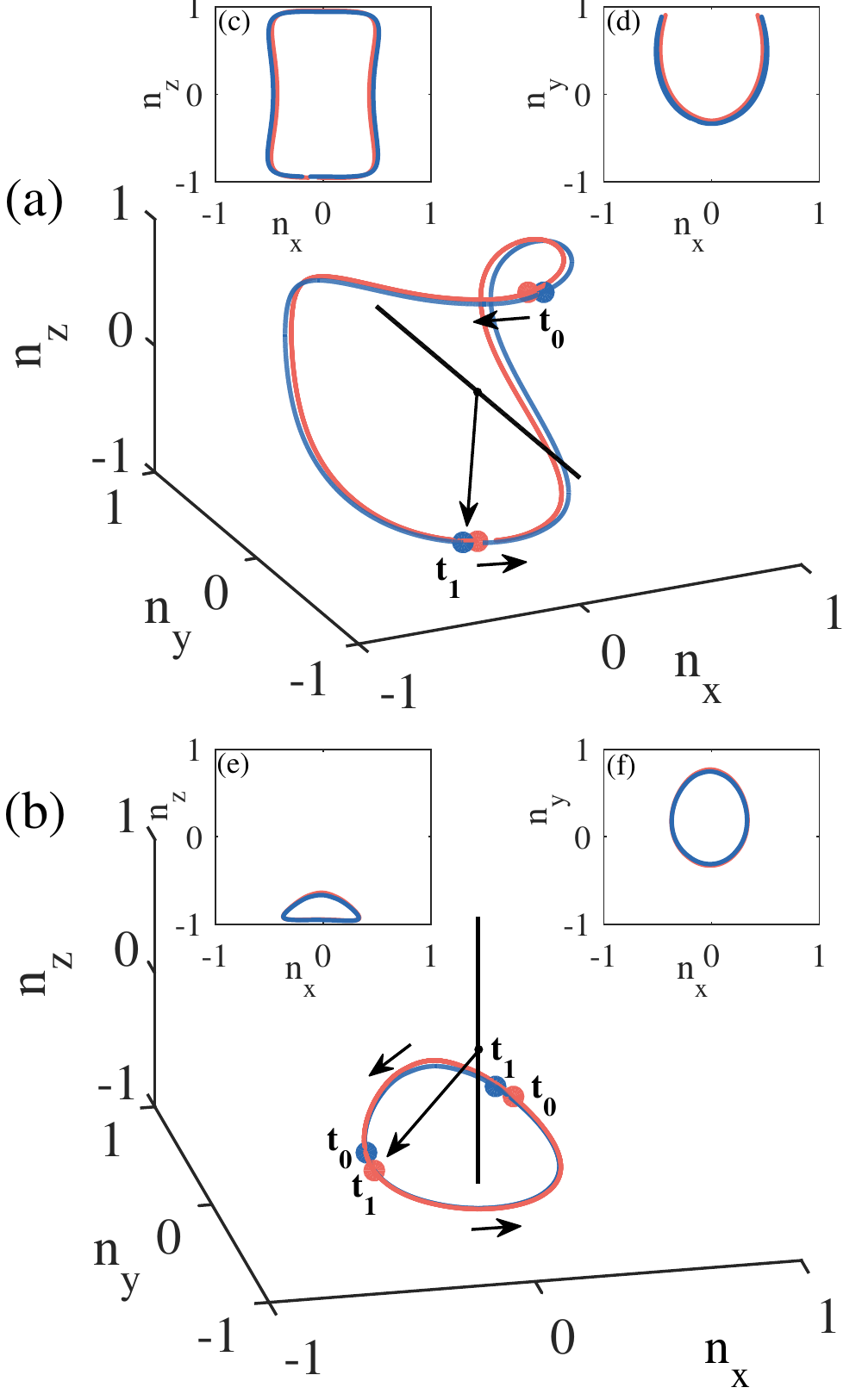}
\caption{ {\bf Trajectories of the magnetization unit vectors}.
(a) Trajectories (red and blue for the two FMIs, respectively) in the 3D
magnetization space corresponding to Fig.~\ref{fig:series2}(a).
The black arrow denotes the trajectory evolution direction. The insets (c,d)
correspond to the projections of the trajectory on the $n_x-n_z$ and
$n_x-n_y$ planes, respectively. (b) Trajectories corresponding to
Fig.~\ref{fig:series2}(b).}
\label{fig:trajectory}
\end{figure}

\begin{figure}
\centering
\includegraphics[width=\linewidth]{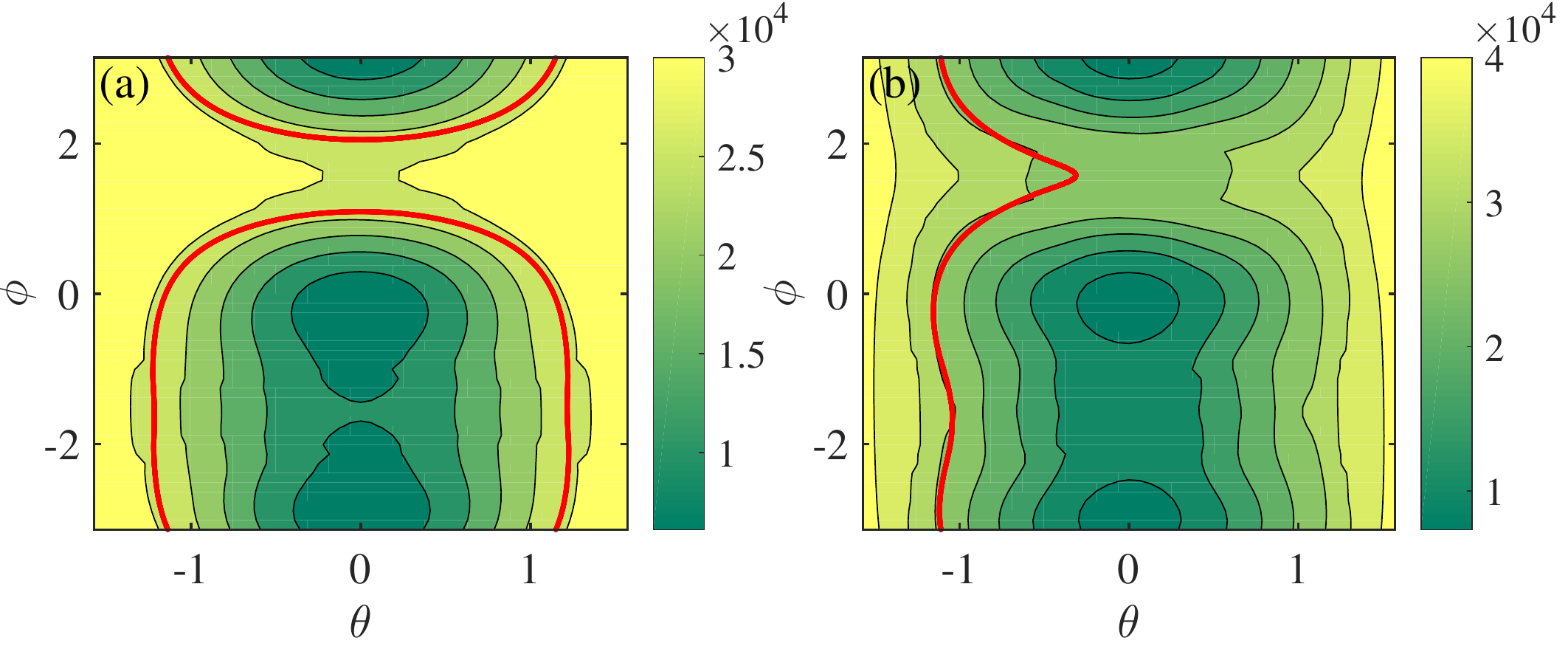}
\caption{ {\bf Magnetization trajectories in the spherical coordinate
with respect to energy distribution}. The red curve is the trajectory of
the magnetization vector. Darker background color indicates lower energy
value. The applied voltage is $55$ meV for (a) and $95$ meV for (b), whereas
the electron energy is $60$ meV for (a) and $30$ meV for (b).}
\label{fig:traj-Edist}
\end{figure}

To examine more closely the different phase and anti-phase locking behaviors
in Fig.~\ref{fig:series2}, we plot the 3D trajectories of the magnetization
unit vector. Figures~\ref{fig:trajectory}(a,b) correspond to the cases
in Figs.~\ref{fig:series2}(a,b), respectively. The red and blue trajectories
are for the two FMIs, and the red and blue dots denote the positions of
the magnetization vector at certain time. For case (a), the trajectories
almost coincide with each other and the magnetization vectors (red and blue
dots) are at the same location for any time, and the frequency of the $y$
component is twice those of the $n_x$ and $n_z$ components, as illustrated in
insets (c,d). For case (b), the trajectories are close to each other but the
magnetization vectors are dominated by the $z$ component and have opposite
phases at the time instants $t_0$ and $t_1$. In this case, the frequencies of
the three components are the same.  We also plot the trajectory of one FMI in
the spherical coordinate, as shown in Fig.~\ref{fig:traj-Edist}, where the
components of the magnetization vector are $n_x = \cos \theta \cos \phi$,
$n_y = \cos \theta \sin \phi$, and $n_z = \sin \theta$.
The spherical coordinate trajectory in Fig.~\ref{fig:traj-Edist}(a)
corresponds to the case in Fig.~\ref{fig:trajectory}(a), where the
magnetization vector circulates about the minimum energy point. In
Fig.~\ref{fig:traj-Edist}(b), the trajectory is along the edge.

\begin{figure}
\centering
\includegraphics[width=\linewidth]{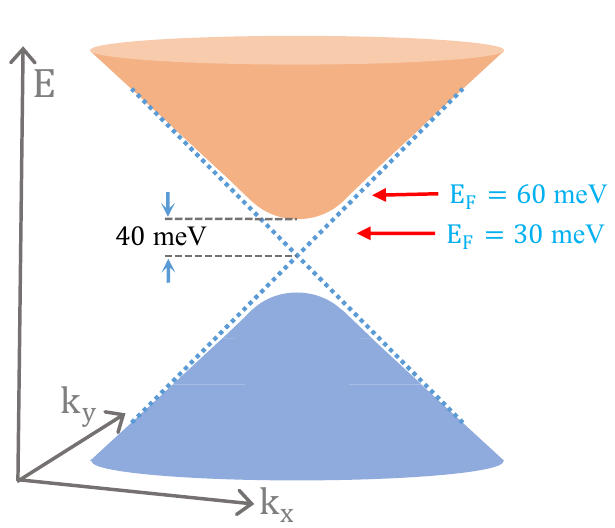}
\caption{ {\bf Electron energy band structure} for the case where the
magnetization vector is in the $z$ direction.}
\label{fig:massgap}
\end{figure}

\begin{figure}
\centering
\includegraphics[width=\linewidth]{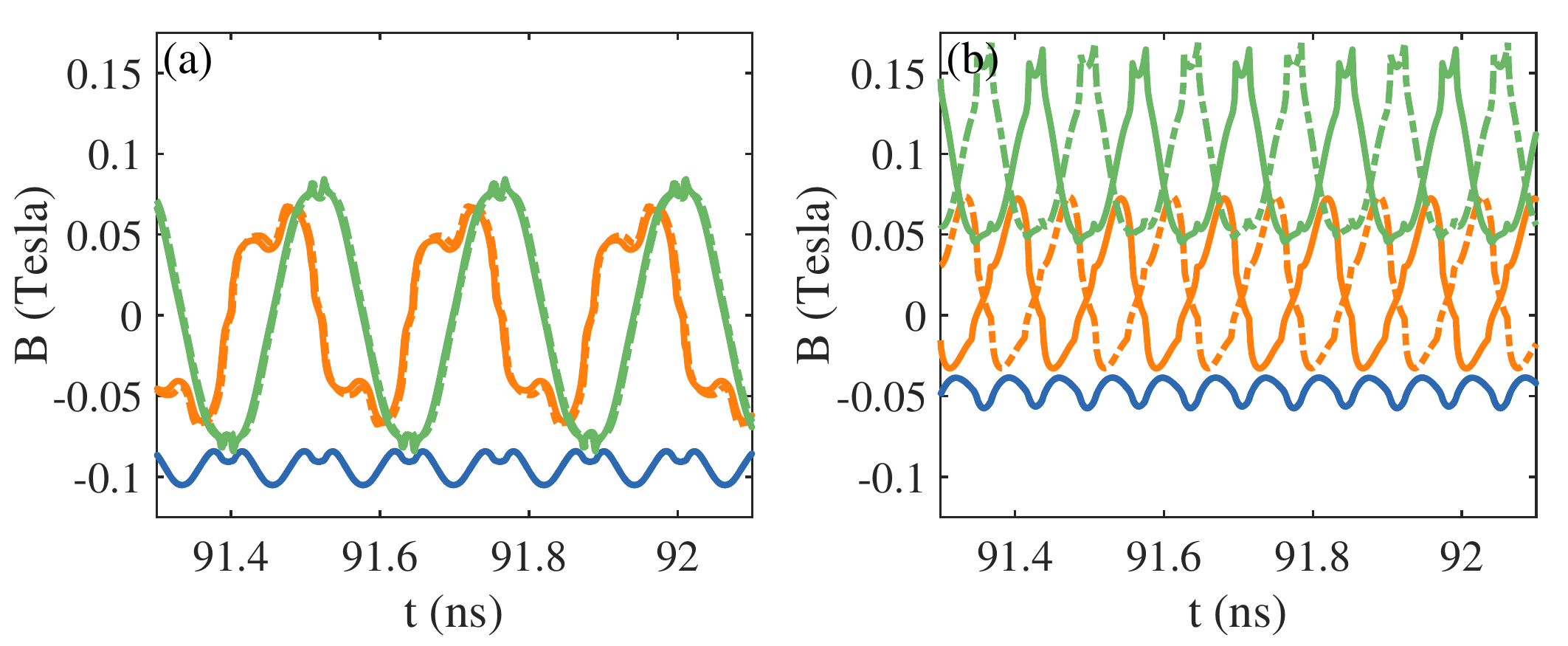}
\caption{ {\bf Effective magnetic field by spin}. The red, blue and green
curves denote the effective magnetic field in the $x$, $y$ and $z$ directions,
respectively. The solid and dashed curves are for the two FMIs. The two cases
are: (a) $E = 60$ meV (above the bottom of the upper band, the upper
horizontal red arrow in Fig.~\ref{fig:massgap}) and (b) $E = 30$ meV (in
the gap, the lower horizontal red arrow in Fig.~\ref{fig:massgap}).}
\label{fig:spinfield}
\end{figure}

The difference in the trajectories is closely related to the relative value
of the electron and exchange coupling energies. As illustrated in
Fig.~\ref{fig:massgap}, when the magnetization vector is along the $z$
direction, the exchange coupling energy is $40$ meV, opening up a gap
in the energy band. For the case where the electron energy is above the
bottom of the upper band (e.g., for energy value of $60$ meV), the
in-plane electron spin component is large, especially along the $y$ axis.
The value of the out-of-plane spin component is limited by the small
exchange coupling energy in comparison with the electron energy. This
has been confirmed by the effective magnetic field value from the average
spin, as shown in Fig.~\ref{fig:spinfield}(a). It can be seen that the
absolute value of the effective magnetic field is stable and large along
the $y$ axis, whereas the $z$ component exhibits large oscillations.
The effective magnetic field by anisotropy makes the magnetization vector
circulate about the $y$ axis. When the electron energy is decreased to, say,
$30$ meV (an energy value inside the gap), the electron will experience a
strong barrier if the $z$ component of the magnetization vector is non-zero,
leading to a large out-of-plane spin component that in turn acts as an
effective magnetic field in the $z$ direction. As a result, the total
effective magnetic field is large in the $z$ direction, causing the
magnetization vector to precess dominantly about the $z$ axis. This
picture is confirmed by the effective magnetic field value experienced by
the electron, as shown in Fig.~\ref{fig:spinfield}(b), where the $z$
component of the field is quite large.

When there is coupling between the two oscillators by the electron
current in the TI, the magnetization vector will be mostly in-plane. In
this case, the anti-damping torque will assume a relatively small value
if there is anti-phase locking between the two magnetization vectors.
As a result, in-phase locking will induce large fluctuations in both $y$
and $z$ components, as can be seen from Fig.~\ref{fig:spinfield}(a),
where the lower values of the $y$ component correspond to a large
absolute value in the $z$ direction. On the contrary, if the out-of-plane
spin component dominates, an anti-damping torque will arise, reducing
the total current fluctuations.

\subsubsection{Effect of quantum interference on phase locking}
\label{subsubsec:WQI}

Having uncovered the phenomena of phase and anti-phase locking in a pair
of FMIs coupled by the spin polarized surface current of the TI, we address
the issue of quantum interference and investigate its effect on the phase
locking dynamics. To take into account quantum interference, we treat the
two FMIs as a single tunneling system and calculate the probability of
quantum tunneling through the whole system.

\begin{figure}
\centering
\includegraphics[width=\linewidth]{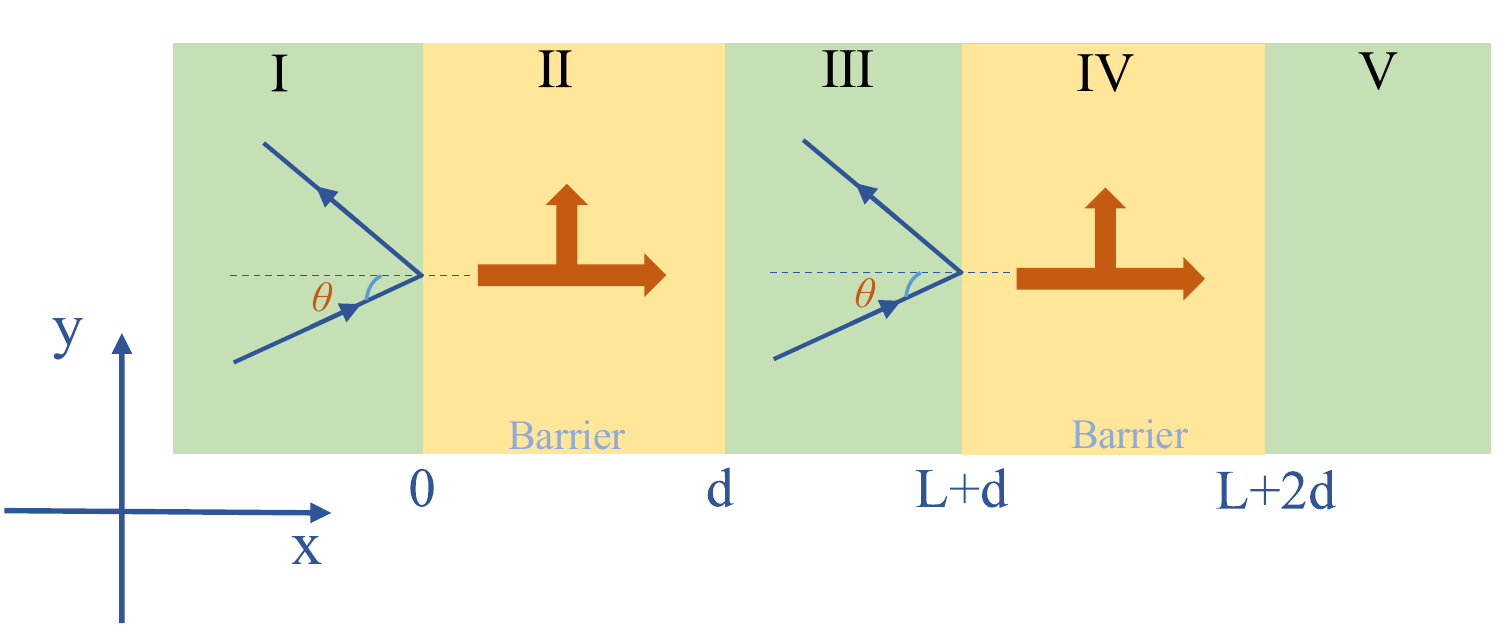}
\caption{ {\bf A schematic illustration of distinct quantum transport
regions for calculating the effective coupling field}. The distance
between the two FMIs is $L=d=40$ nm.}
\label{fig:schematic_2}
\end{figure}

Consider a surface electron in the TI moving toward the interface region.
As shown in Fig.~\ref{fig:schematic_2}, there are five subregions of
interest: (I) the ``free'' region to the left of the interface between the
first FMI and TI, (II) the interface region itself, (III) the region
between the two interfaces, (IV) the second interface region, and (V) the
``free'' region to the right of the region IV. Let $\theta$ be the incident
angle of the electron from region I to region II. Solving the Dirac equation
in each subregion, we obtain the spinor wavefunctions in the five regions, as
listed in Appendix~\ref{App_D_QIF}. Matching the wavefunctions at the
boundaries, we obtain all the coefficients and hence the wavefunction in
the whole 2D space. The average spin polarization in each subregion and the
corresponding effective magnetic field can then be calculated, as in
Eqs.~(\ref{Current_Density})-(\ref{eq:LLGEq}).

\begin{figure}
\centering
\includegraphics[width=\linewidth]{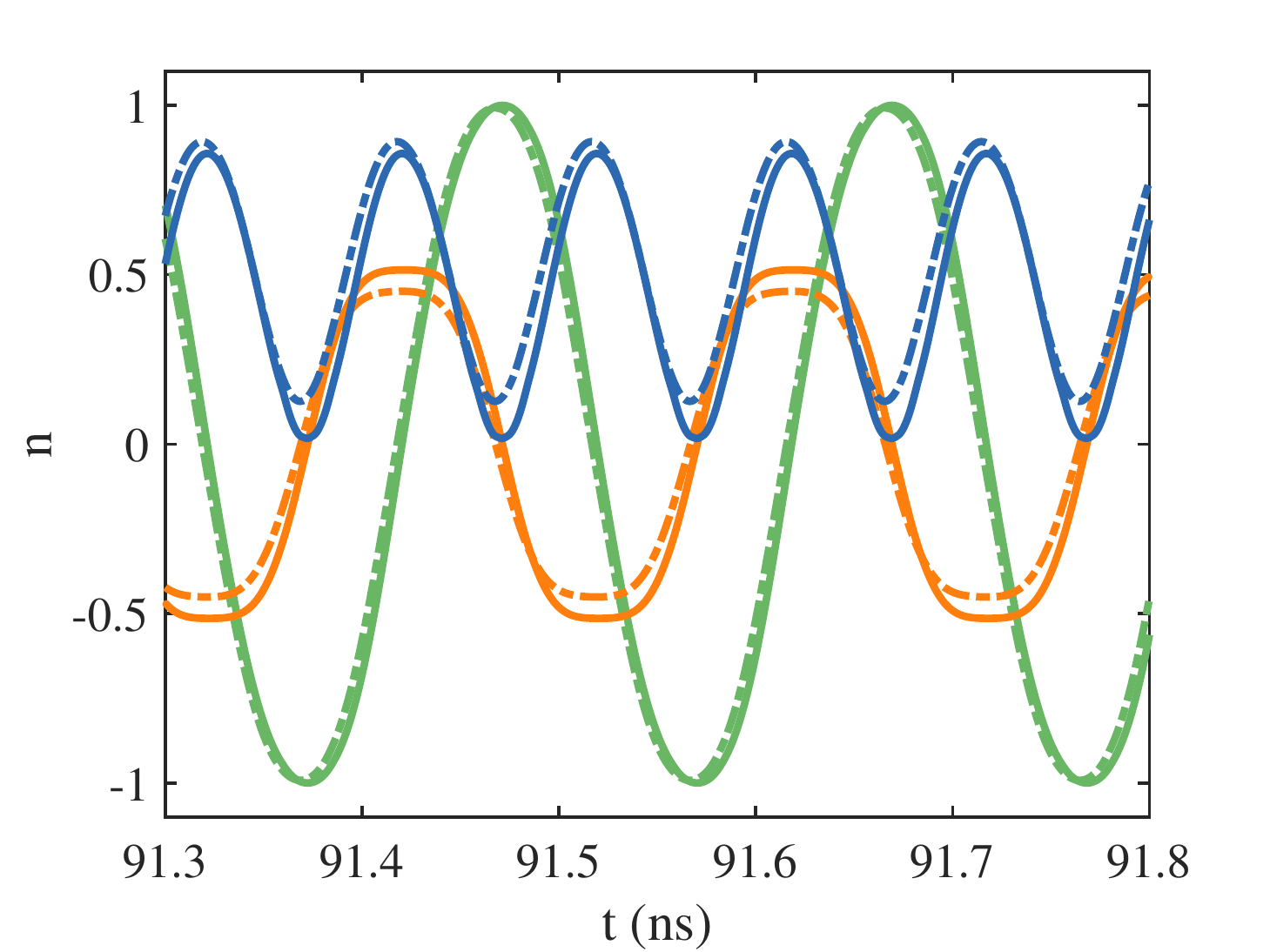}
\caption{ {\bf Phase locking between the two coupled FMIs in the presence
of quantum interference}. The applied voltage is $50$ mV and the electron
energy is $100$ meV. The red, blue and green curves denote $n_x$, $n_y$
and $n_z$, respectively. The solid and dashed curves are for the FMI in
region II and IV, respectively.}
\label{fig:coherent}
\end{figure}

Figure~\ref{fig:coherent} shows the representative magnetization dynamics
of the two FMIs when quantum interference is taken into account (for
$V_0 = 50$ mV and $E = 100$ meV). The three components of the magnetization
vector are represented by different colors, and the solid and dashed curves 
are for the first and second FMI, respectively. The magnetization vectors
exhibit oscillations and there is phase locking. We vary the external voltage
and the electron energy and also change the anisotropy value. In all cases, 
persistent phase locking is observed.

\section{Conclusion} \label{sec:conclusion}

To summarize, motivated by the need for nanoscale oscillators for
developing potential unconventional computing paradigms, we have studied
the oscillatory dynamics and phase locking of a pair of FMI oscillators
coupled through the spin-polarized current on the surface of an TI.
The dynamics of the whole system are of the hybrid type~\cite{WXL:2018}: a
combination of classical nonlinear and relativistic quantum dynamics, with
the following underlying physics. For each heterostructure interface between
the FMI and TI, there is an average spin of the electron flow, which can
be solved via the transmission through the interface. The average spin acts
as an effective field which, when combining with the magnetic anisotropy
of the FMI, leads to self oscillations in the magnetization of each individual,
uncoupled FMI. The self oscillations in turn modulate the electron 
transmission periodically, making the surface current of the TI time varying. 
The AC current generates the coupling between the two FMIs. As a result, 
stable phase or anti-phase locking between the two FMIs emerges, regardless 
of whether quantum interference is absent or present. The phase locking
phenomenon is robust as it occurs in wide ranges of the external applied
voltage and electron energy. To our knowledge, this is the first demonstration
of phase locking due to proximity effect induced torques in FMI/TI systems,
justifying further investigation of these systems in terms of their possible
role in serving as the fundamental building block of unconventional computing
paradigms.

Some realistic considerations are the following.
In an experimental setup, if the two FMIs are far from each other (e.g.,
$> 100$ nm), scattering from impurities will destroy the coherence between
the states of the two FMIs. In this case, our non-coherent approach is
applicable. If the two FMIs are close to each other (e.g., within 100 nm),
coherence cannot be ignored, rendering necessary our quantum coherence based
treatment. 

Direct interaction between the two FMIs can also affect the phase locking 
dynamics. One such type is the dipole-dipole interaction~\cite{chen2016phase} 
described by the Hamiltonian
\begin{align}
H_{dip} = -\frac{\mu_0}{4\pi |\boldsymbol{r}|^3} [3(\boldsymbol{m}_1\cdot \hat{\boldsymbol{r}})(\boldsymbol{m}_2\cdot \hat{\boldsymbol{r}}) - \boldsymbol{m}_1 \cdot \boldsymbol{m}_2],
\end{align}
where $\mu_0$ is the vacuum permeability, $\boldsymbol{r}$ is the distance
between the two effective point dipoles, $\hat{\boldsymbol{r}}$ is a unit
vector parallel to the line joining the centers of the two dipoles, and
$\boldsymbol{m}_{1,2} = \boldsymbol{M}_{1,2}\cdot V$ with $V$ being the volume
of the FMI stripe. For one FMI, the effective magnetic field from the second FMI
is $\boldsymbol{B}_{1,2} = -\partial H/\partial\boldsymbol{M}_{1,2}$.
Setting the two magnetization vectors in the same direction (the configuration
of minimum energy) and using our simulation parameter setting, we estimate
the energy density to be $H/V \approx 2\times 10^4$ erg/cm$^3$ for
$|\boldsymbol{r}|=20$ nm. This is about one order of magnitude smaller than
the anisotropy coefficient in the $z$ direction
($K_z = 2.5\times 10^{5}$ erg/cm$^3$). Insofar as the edge distance between
the two FMIs is larger than $20$ nm, dipole-dipole interaction can be
neglected. If the two FMIs are too close to each other, the dipole-dipole
energy can be comparable to the system energy, which cannot be ignored.
The effect of dipole-dipole interaction on phase locking is a topic that
warrants further study.

\section*{Acknowledgement}

We thank Prof.~L.~ Huang for helpful discussions. We would like to acknowledge
support from the Vannevar Bush Faculty Fellowship program sponsored by the
Basic Research Office of the Assistant Secretary of Defense for Research and
Engineering and funded by the Office of Naval Research through Grant
No.~N00014-16-1-2828.

\appendix

\section{Experimental realizability of phase locking in the FMI/TI heterostructure}
\label{App_A_exp}

A typical TI/FMI system is the Bi$_2$Se$_3$/YIG (yttrium iron garnet)
heterostructure~\cite{lang2014proximity,wang2016surface,fanchiang2018strongly},
where the effect of exchange interaction between the FMI and the TI surface 
states on the magnetization dynamics of YIG has been studied 
recently~\cite{fanchiang2018strongly}. Among the different types of 
anisotropy in the YIG thin film, shape anisotropy is dominant. In 
particular, the hard axis is perpendicular to the plane of the film ($z$ 
direction in our study) and the associated anisotropy coefficient is on the 
order of $K_z \sim 10^5$ erg/cm$^3$ when the thickness $d$ of the film is in 
the range from several nm to tens of 
nm~\cite{fanchiang2018strongly,wang2016surface}. The magnetocrystalline 
anisotropy coefficient is smaller than that of the shape anisotropy, which 
is about $K \sim 2.5\times 10^4$ erg/cm$^3$ and can produce a hard axis in 
the plane favoring magnetization along the $\langle 111 \rangle$ 
axis~\cite{wang2016surface}. The typical value of the magnetization is on the 
order of 1000 Oe~\cite{fanchiang2018strongly,wang2016surface}.
Another widely studied heterostructure Bi$_2$Se$_3$/EuS~\cite{wei2013exchange,
li2015proximity,katmis2016high,lv2018large}, where progress on 
magnetoresistance and current-induced magnetization switching has been recently 
reported~\cite{lv2018large}. For EuS, the hard axis anisotropy is in the
range~\cite{li2015proximity,katmis2016high} of $10^4 - 10^5$ erg/cm$^3$, and 
the value of magnetization~\cite{katmis2016high} is also on the order of 
$1000$ Oe. The anisotropy value used in our study,
$K_z=2.5\times 10^5$ erg/cm$^3$, is comparable to those of the two materials.
It is thus potentially feasible to realize auto-oscillation and then phase 
locking using the Bi$_2$Se$_3$/YIG or the Bi$_2$Se$_3$/EuS heterostructure.

\begin{figure}[h]
\begin{center}
\includegraphics[width=\linewidth]{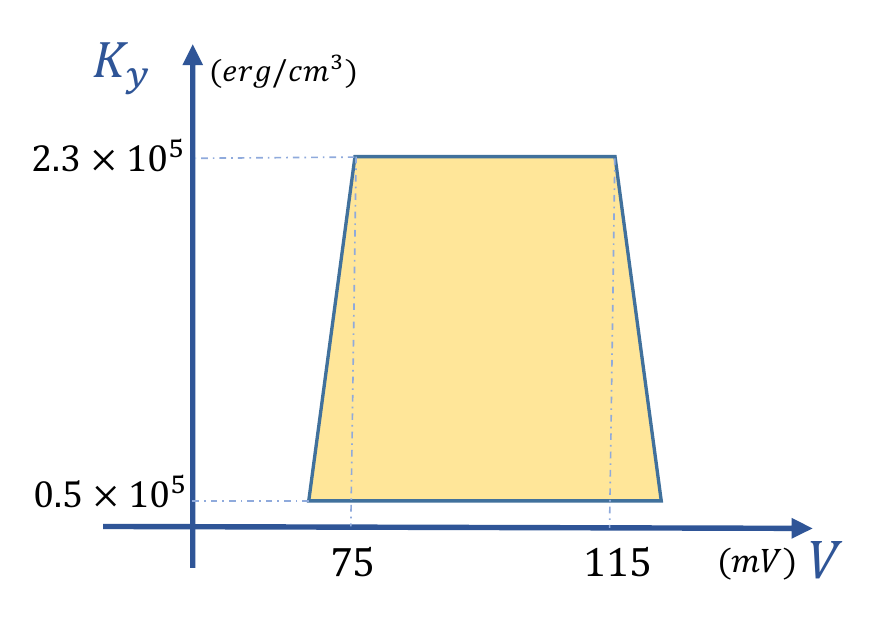}
\end{center}
\caption{ {\bf Typicality of phase locking in a representative parameter
space}. For fixed $K_z = 2.5\times 10^5$ erg/cm$^3$ and Fermi energy
$E = 100$ meV, the region in the parameter plane $(K_y,V)$ for phase locking
(the yellow area).}
\label{fig:ani_range}
\end{figure}

While we have used specific anisotropy values to illustrate the phase locking
phenomenon in the main text, these values can be tuned in certain ranges
without losing phase locking, making it possible to match the values to those
of real materials. Specifically, if the value of anisotropy in the $z$
direction is fixed, phase locking can be achieved by varying the value of the
anisotropy in the $y$ direction and the applied voltage in a wide range. For
example, say the anisotropy coefficient in the $z$ direction is
$K_z = 2.5\times 10^5$ erg/cm$^3$. We find that phase locking can be achieved
when the $K_y$ value varies in a wide range and the voltage $V$ can also be
chosen from a range, as shown in Fig.~\ref{fig:ani_range}, where the electron
Fermi energy is $E = 100$ meV. The yellow area in Fig.~\ref{fig:ani_range}
indicates the approximate parameter region for phase locking. We find,
however, that it is necessary to restrict the anisotropy difference of the two
nanocontacts to being less than $10\%$ of the $K_z$ value in order to realize 
phase locking.

The value of $K_z$ can also be tunned without losing phase locking,
which can be seen from the following heuristic analysis based on the LLG
equation. Note that the LLG equation can be reduced to the
Landau-Lifshitz (LL) equation by substituting the cross product of
$\boldsymbol{n}$ on the left side
\begin{align}
\boldsymbol{n} \times \frac{d\boldsymbol{n}}{dt} = -\gamma \boldsymbol{n}\times (\boldsymbol{n}\times \boldsymbol{B}_{eff}) - \alpha \frac{d\boldsymbol{n}}{dt},
\end{align}
into the LLG equation. We have
\begin{align}
\frac{d\boldsymbol{n}}{dt} = -\frac{\gamma}{1+\alpha^2}\boldsymbol{n}\times \boldsymbol{B}_{eff} -\frac{\gamma \alpha}{1+\alpha^2}\boldsymbol{n}\times (\boldsymbol{n}\times \boldsymbol{B}_{eff}).
\end{align}
In general, increasing the effective magnetic field is equivalent to
decreasing the time period of magnetization oscillations. Specifically,
we have $\boldsymbol{B}_{eff} = \boldsymbol{B}_{spin} + \boldsymbol{B}_{an}$,
where the spin induced effective magnetic field is proportional to the applied
voltage: $\boldsymbol{B}_{spin} \sim V$, and the anisotropic magnetic field
is proportional to the anisotropy coefficient: $\boldsymbol{B}_{an} \sim K$.
When the value of the material anisotropy is altered, say, within one order
of magnitude, it is just necessary to change the voltage by the same amount
to ensure that the value of $\boldsymbol{B}_{eff}$ changes by a proper amount.
With such parameter changes, while the oscillation period or the characteristic
time scale underlying the magnetization dynamics is changed, phase locking
is maintained.

\section{Iterative solution method for solving the coupled magnetization 
dynamics}
\label{App_B_fig}

\begin{figure}[h]
\begin{center}
\includegraphics[width=\linewidth]{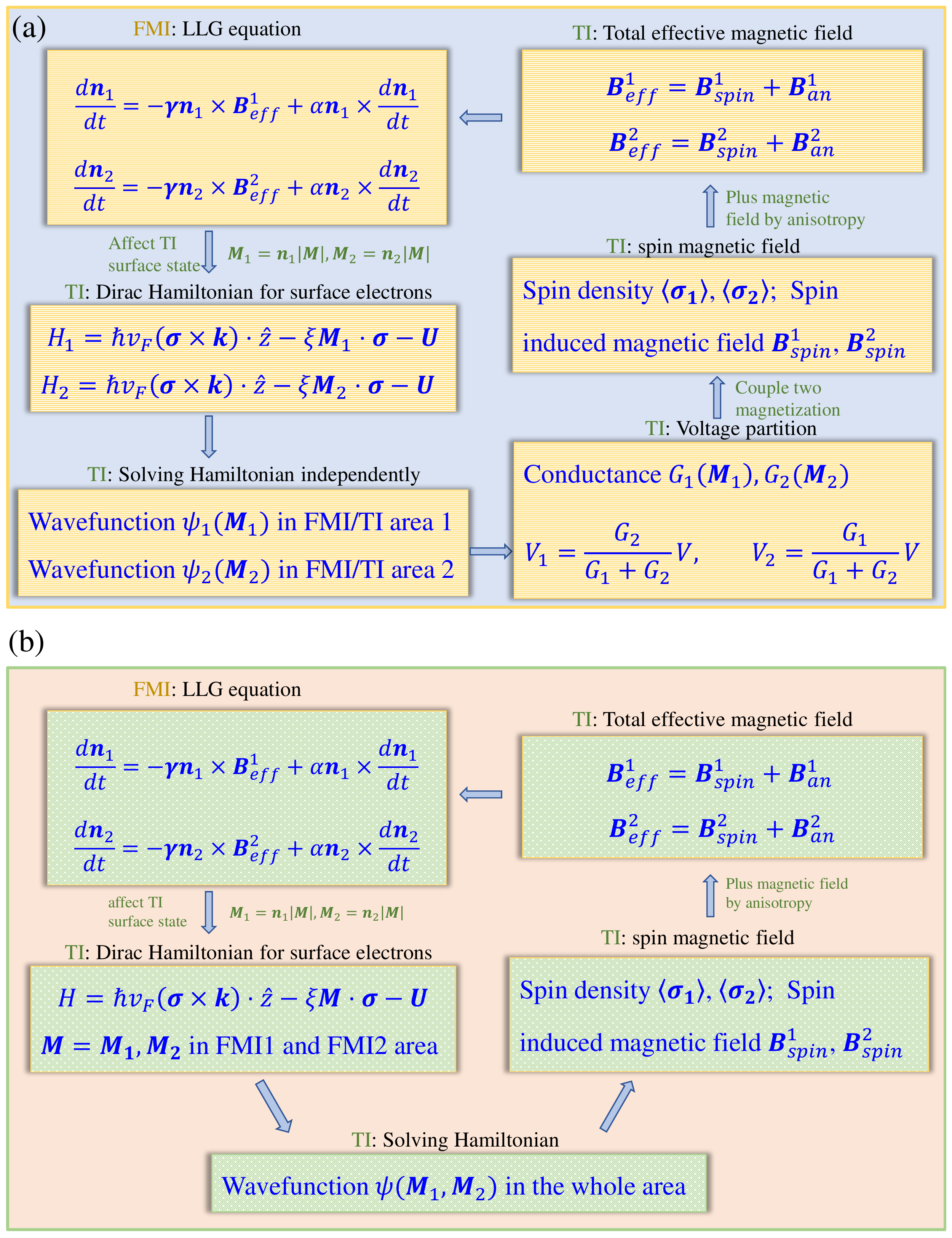}
\end{center}
\caption{ {\bf Iterative procedure for solving the coupled LLG equations for  
the magnetization dynamics}. The two FMIs are coupled by the surface current
of the TI. (a,b) Without and with quantum interference, respectively.} 
\label{fig:Iterative_procedure}
\end{figure}

Figure~\ref{fig:Iterative_procedure} presents a flow chart detailing our
iterative procedure for solving the coupled LLG equations for the 
magnetization dynamics for the two cases where quantum interference is
absent and present, respectively. All the quantities have been defined in 
Sec.~\ref{sec:model}. 

\section{Electron spin density calculation} \label{App_C_Iteration}

\subsection{Proof of an equality}

We prove the equality
\begin{displaymath}
T_{\boldsymbol{M}}(E_F,\theta)\cos\theta = -\psi^{\dagger} \sigma_y \psi.
\end{displaymath}

For one FMI/TI heterostructure (regions \uppercase\expandafter{\romannumeral1},
\uppercase\expandafter{\romannumeral2} and
\uppercase\expandafter{\romannumeral3} in Fig.~\ref{fig:schematic_2}), 
the wavefunctions in the incident area, heterostructure interface, and 
the transmitted regions are~\cite{Yokoyama:2011}, respectively, 
\begin{eqnarray}
\nonumber
\psi_1(x\le0) & = & \frac{1}{\sqrt{2}}
\left( \begin{array}{c}
ie^{-i\theta} \nonumber \\
1
\end{array} \right)
e^{ik_Fx\cos\theta} \\ \nonumber
& + & \frac{r}{\sqrt{2}}
\left( \begin{array}{c}
-ie^{i\theta}\\
1
\end{array} \right)
e^{-ik_Fx\cos\theta} 
\end{eqnarray}

\begin{eqnarray} 
\nonumber
& & \psi_2(0< x< d) = A \cdot \left( \begin{array}{c}
\hbar v_F (\tilde{k}_{y}+i\tilde{k}_{x})\\
E+U-\xi M_{z}
\end{array} \right)
e^{i(\tilde{k}_{x} + \xi M_{y}/\hbar v_F)x} \\ \label{App_Eq:heterostructure}
& & \ \ \ \ \ + B \cdot \left( \begin{array}{c}
\hbar v_F (\tilde{k}_{y}-i\tilde{k}_{x}) \\
E+U-\xi M_{z}
\end{array} \right)
e^{i(-\tilde{k}_{x} + \xi M_{y}/\hbar v_F)x}
\end{eqnarray}

\begin{equation} \label{App_Eq:transmitted}
\psi_3(x\ge d) = \frac{t}{\sqrt{2}}
\left( \begin{array}{c}
ie^{-i\theta} \\
1
\end{array} \right)
e^{ik_Fx\cos\theta}
\end{equation}
where $r$ and $t$ are the reflection and transmission coefficients,
$E+U = \hbar v_F k_F$, $k_x = k_F \cos\theta$, $k_y = k_F\sin\theta$,
$\hbar v_F \tilde{k}_x=\sqrt{(E+U)^2-(\xi M_z)^2-(\hbar v_F \tilde{k}_y)^2}$,
and $\hbar v_F \tilde{k}_y = \hbar v_F k_y - \xi M_x$. For convenience, we
denote $A = \frac{a}{\sqrt{2(E+U)(E+U-\xi M_z)}}$ and
$B = \frac{b}{\sqrt{2(E+U)(E+U-\xi M_z)}}$ with $a$ and $b$ being the
corresponding coefficients. Matching the wavefunctions at the boundary
between different regions, we get the corresponding coefficients, $r$,
$a$, $b$, and $t$. For simplicity, in the wavefunction expressions, the
spatial normalization factor $1/\sqrt{LW}$ is not included, where $L$ and
$W$ are the size parameters of the 2D device. For an electron, the $y$
component of the spin in the transmitted
region~\uppercase\expandafter{\romannumeral3} can be obtained via the
wavefunction [Eq.~(\ref{App_Eq:transmitted})] average
\begin{align}
\psi_3^{\dagger}\sigma_y\psi_3 = -|t|^2\cos\theta,
\end{align}
The transmission probability is $T(E, \theta) = |t|^2$. We have
$T\cos\theta = -\psi_3^{\dagger}\sigma_y \psi_3$, which is the relation
between electron transmission in the $x$ direction and the $y$ component of the
spin in the transmitted region (free region), i.e., spin-momentum locking. In
fact, this relation is also valid in the FMI/TI heterostructure region, i.e.,
$\psi_2^{\dagger} \sigma_y \psi_2=\psi_3^{\dagger} \sigma_y \psi_3$ holds
independent of position $x$, which can be proved, as follows. We write down the
spin value in terms of the wavefunction [Eq.~(\ref{App_Eq:heterostructure})]
in the heterostructure region at position $x$ as
\begin{align} \label{Eq:sigmay_dist}
\psi_2^{\dagger} \sigma_y \psi_2 &= -2 \beta \cdot \Big[ |A|^2 \Im (\alpha_1) \exp(-2\Im(\tilde{k}_x)x) \\
&+ |B|^2 \Im(\alpha_2) \exp(2\Im(\tilde{k}_x)x)\Big] \nonumber \\
&+2\beta\cdot \Re\Big[ iA^*B\cdot (\alpha_2 - \alpha_1^*) \exp(-2i\Re(\tilde{k}_x )x)\Big] \nonumber
\end{align}
where $\alpha_1 \equiv \hbar v_F (\tilde{k}_{y}+i\tilde{k}_{x})$,
$\alpha_2 \equiv \hbar v_F (\tilde{k}_{y}-i\tilde{k}_{x})$,
$\beta \equiv E + U - \xi M_z$, $\Re(\cdot)$ and $\Im(\cdot)$ represent
the real and imaginary parts, respectively. The expression in
Eq.~(\ref{Eq:sigmay_dist}) can be further simplified. Note that
$\tilde{k}_y$ is real and $\tilde{k}_x$ can be real or purely imaginary
depending on the quantities $M_z$ and $k_y$
[c.f., $\hbar v_F\tilde{k}_x=\sqrt{(E+U)^2-(\xi M_z)^2-(\hbar v_F \tilde{k}_y)^2}$].
To evaluate the value of $\psi_2^{\dagger} \sigma_y \psi_2$, we first assume
$\tilde{k}_x$ is real. In this case, the second part on the right-hand side
of Eq.~(\ref{Eq:sigmay_dist}) is zero, and
$\exp(-2\Im(\tilde{k}_x)x)=\exp(2\Im(\tilde{k}_x)x)=1$. We thus have
\begin{align}
\psi_2^{\dagger} \sigma_y \psi_2 = -2\beta\hbar v_F \tilde{k}_x(|A|^2 - |B|^2),
\end{align}
which is independent of the position $x$. We then consider the case where
$\tilde{k}_x$ is purely imaginary. In this case, the first part on the 
right-hand side of Eq.~(\ref{Eq:sigmay_dist}) is zero, because
$\Im(\alpha_1) = \Im(\alpha_2) = 0$. Using $\exp(-2i\Re(\tilde{k}_x x))=1$,
we have
\begin{align}
\psi_2^{\dagger} \sigma_y \psi_2 = 4\beta \Re(A^*B \tilde{k}_x).
\end{align}
which is also independent of position $x$. Utilizing wavefunction
matching at the boundary, we see that the equality
$T(E, \theta) \cos\theta = -\psi^{\dagger}_3 \sigma_y \psi_3 = 
-\psi^{\dagger}_2 \sigma_y \psi_2$ holds in the heterostructure region.

\subsection{Calculation of the spin density}

The current density along the $x$ direction [Eq.~(\ref{Current_Density})] can 
be written in terms of the spin average, i.e.,
\begin{align} \label{Current_Density_spin}
J_x &= \frac{V_1 G_1}{L_w}= -\frac{Ee^2 V_1}{2\pi^2 \hbar^2 v_F}
\int^{\frac{\pi}{2}}_{-\frac{\pi}{2}} \psi_2^{\dagger} \sigma_y \psi_2 d\theta.
\end{align}
Utilizing the current definition in the free area:
$\hat{J}=-e \nabla_{\boldsymbol{p}}H = -ev_F(-\hat{\sigma}_y, \hat{\sigma}_x)$,
we get the mean spin density as
\begin{align} \label{Eq:sigma_y_density}
\langle \sigma_y \rangle_1 &= \frac{J_x}{ev_F}= -\frac{Ee V_1}{2\pi^2 \hbar^2 v^2_F }\int^{\frac{\pi}{2}}_{-\frac{\pi}{2}} \psi^{\dagger}_3 \sigma_y \psi_3 d\theta \nonumber\\
&= -\frac{Ee V_1}{2\pi^2 \hbar^2 v^2_F }\int^{\frac{\pi}{2}}_{-\frac{\pi}{2}} \psi^{\dagger}_2 \sigma_y \psi_2 d\theta \nonumber\\
&= -\frac{Ee V_1}{2\pi^2 \hbar^2 v^2_F d}\int^{d}_0\int^{\frac{\pi}{2}}_{-\frac{\pi}{2}} \psi^{\dagger}_2 \sigma_y \psi_2 d\theta dx.
\end{align}
For an incident electron with certain angle and energy, the spin value at
each point $x$ in the heterostructure is given by
$(\psi_2^{\dagger} \sigma_x \psi_2, \psi_2^{\dagger} 
\sigma_y \psi_2,\psi_2^{\dagger} \sigma_z \psi_2)$. We perform the angle and
position averages for $\sigma_x$ and $\sigma_z$ as for $\sigma_y$, and
multiply the factor related to the electron density, as in
Eq.~(\ref{Eq:sigma_y_density}). The $x$ and $z$ components of the spin
density in the heterostructure region can then be obtained by substituting
$\psi_2^{\dagger} \sigma_y \psi_2$ with $\psi_2^{\dagger} \sigma_x \psi_2$
and $\psi_2^{\dagger} \sigma_z \psi_2$ in Eq.~(\ref{Eq:sigma_y_density}):
\begin{align}
\langle \sigma_x \rangle_1 = -\frac{Ee V_1}{2\pi^2 \hbar^2 v^2_F d}\int^{d}_0\int^{\frac{\pi}{2}}_{-\frac{\pi}{2}} \psi^{\dagger} \sigma_x \psi d\theta dx,\\
\langle \sigma_z \rangle_1 = -\frac{Ee V_1}{2\pi^2 \hbar^2 v^2_F d}\int^{d}_0\int^{\frac{\pi}{2}}_{-\frac{\pi}{2}} \psi^{\dagger} \sigma_z \psi d\theta dx.
\end{align}

\subsection{Spin density in a 2D Rashba plane}

The general Rashba Hamiltonian with exchange interaction has the 
form~\cite{manchon2008theory,manchon2009theory,miron2010current} 
\begin{align}
H = \frac{\hbar^2}{2m}k^2 + \alpha(\boldsymbol{\sigma}\times\boldsymbol{k})\cdot \hat{z} - \xi \boldsymbol{M}\cdot \boldsymbol{\sigma},
\end{align}
where $\hbar \boldsymbol{k} = -i \hbar (\partial_x, \partial_y, 0)$ is the
two-dimensional electron momentum operator, $\alpha$ parametrizes the
spin-orbit coupling, $\xi$ is the exchange coupling strength between
conduction electron and magnetization. We can write down the wavefunctions
in the incident, heterostructure and transmitted areas, and match the
wavefunctions at the boundary to get the corresponding coefficients. For 
example, for the transmitted wave in the following 
form~\cite{inoue2003diffuse} (which is needed for calculating the spin density):
\begin{align}
\phi_{t} = \frac{t}{\sqrt{2}}\exp(i\boldsymbol{k}\cdot\boldsymbol{r})\cdot
\left( \begin{array}{c}
i\cdot s \exp(-i\theta) \\
1
\end{array} \right)
\end{align}
where $t$ is the transmission coefficient, $s=\pm1$, and
$\exp(-i\theta) = (k_x - ik_y)/\sqrt{k_x^2 + k_y^2}$. The current operator
in the $x$ direction in the transmitted area is defined as
\begin{align}
\hat{J}_x=-e \nabla_{p_x}H=-e(\frac{p_x}{m}-\frac{\alpha}{\hbar}\hat{\sigma}_y).
\end{align}
For an electron, the current with incident angle $\theta$ is
\begin{align}
j_x &= \phi^{\dagger}_t \hat{J}_x \phi_t = -e\Big[\frac{p_x}{m}|t|^2 + s\cdot \frac{\alpha}{\hbar}|t|^2\cos\theta\Big] \nonumber \\
&=-e \Big[\frac{\hbar k}{m} + \frac{s\alpha}{\hbar} \Big]|t|^2 \cos\theta
\end{align}
The angle averaged current is
\begin{align}
j^{ave}_x = \frac{-e}{\pi} \cdot \Big[\frac{\hbar k}{m} + \frac{s\alpha}{\hbar} \Big]\int^{\frac{\pi}{2}}_{-\frac{\pi}{2}}
T_{\boldsymbol{M}}(E, \theta) \cos \theta d\theta.
\end{align}
From the classical Landauer-Buttiker 
formalism~\cite{WPC:2010,SDK:2014,Datta:book}, we have the conductance as
\begin{align}
G = \frac{Ee^2 L_w}{2\pi^2 \hbar^2 v_F}\int^{\frac{\pi}{2}}_{-\frac{\pi}{2}}
T_{\boldsymbol{M}}(E, \theta) \cos \theta d\theta.
\end{align}
The current density is
\begin{align}
J_x = \frac{V_1G_1}{L_w}=\frac{Ee^2 V_1}{2\pi \hbar^2 v_F}\cdot \frac{1}{\pi}\int^{\frac{\pi}{2}}_{-\frac{\pi}{2}}
T_{\boldsymbol{M}}(E, \theta) \cos \theta d\theta.
\end{align}
The incident electron density can then be expressed as
\begin{align}
n = \frac{J_x}{j^{ave}_x} = -\frac{EeV_1}{2\pi \hbar^2 v_F}\cdot \frac{1}{\frac{\hbar k}{m} + \frac{s\alpha}{\hbar}}.
\end{align}
Once the current density is obtained, we can calculate the spin average over
different incident angles and positions. For example, the $\sigma_y$ component
can be written as
\begin{align}
\langle \sigma_y \rangle_1 &= n\cdot \frac{1}{\pi d}\int^{d}_0\int^{\frac{\pi}{2}}_{-\frac{\pi}{2}} \psi^{\dagger} \sigma_y \psi d\theta dx \nonumber \\
&= -\frac{Ee V_1}{2\pi^2 \hbar^2 v_F d}\cdot \frac{1}{\frac{\hbar k}{m} + \frac{s\alpha}{\hbar}} \int^{d}_0\int^{\frac{\pi}{2}}_{-\frac{\pi}{2}} \psi^{\dagger} \sigma_y \psi d\theta dx,
\end{align}
Taking the limit $m\to \infty$, $\alpha = \hbar v_F $, $s=1$, we get the spin
density for our TI system in Eq.~(\ref{Eq:sigma_y_density}). That is, the
surface states of TI correspond to the $m\to \infty$ limit of the 2D Rashba
Hamiltonian, at which the maximum spin-momentum locking efficiency is achieved.

Another type of 2D Rashba systems can be graphene based heterostructures,
e.g., graphene/Ni(111) and graphene/transition metal dichalcogenide. For
such systems, the electron dynamics are governed by the 2D Dirac-Rashba
Hamiltonian and exhibit significant in-plane spin polarization, which is
perpendicular to electron momentum and proportional to the group 
velocity~\cite{rashba2009graphene}.
Similar to the surface electron states of a topological insulator, an
in-plane voltage induced charge current will produce a spin density along
the perpendicular direction and hence a torque that can be the driven
source for magnetization in adjacent magnetic insulators. Generalizing
the current formalism, we expect similar inter-coupling dynamics 
between electronic transport and magnetization.

\section{Solutions of quantum tunneling of Dirac electrons through
double FMI barriers}\label{App_D_QIF}

The spinor wavefunctions in the five regions in Fig.~\ref{fig:schematic_2} can
be written as
\begin{align} \nonumber
\psi_1(x\le0) = \frac{1}{\sqrt{2}}
\left( \begin{array}{c}
ie^{-i\theta} \nonumber \\
1
\end{array} \right)
e^{ik_Fx\cos\theta}\nonumber \\
+\frac{r}{\sqrt{2}}
\left( \begin{array}{c}
-ie^{i\theta} \nonumber \\
1
\end{array} \right)
e^{-ik_Fx\cos\theta}
\end{align}

\begin{align} \nonumber
\psi_2(0< x\le d) = \frac{1}{\sqrt{2(E+U_1)(E+U_1-M_{z1})}} \nonumber \\
\cdot \Bigg[ a\left( \begin{array}{c}
\hbar v_F (\tilde{k}_{y1}+i\tilde{k}_{x1})\\
E+U_1-M_{z1}
\end{array} \right)
e^{i(\tilde{k}_{x1} + m_{y1}/\hbar v_F)x}\nonumber \\
+b\left( \begin{array}{c}
\hbar v_F (\tilde{k}_{y1}-i\tilde{k}_{x1}) \nonumber \\
E+U_1-M_{z1}
\end{array} \right)
e^{i(-\tilde{k}_{x1} + m_{y1}/\hbar v_F)x}\Bigg]
\end{align}

\begin{align} \nonumber
\psi_3(d<x<L+d) = \frac{c}{\sqrt{2}}
\left( \begin{array}{c}
ie^{-i\theta}\\
1
\end{array} \right)
e^{ik_Fx\cos\theta}\nonumber \\
+\frac{d}{\sqrt{2}}
\left( \begin{array}{c}
-ie^{i\theta} \nonumber \\
1
\end{array} \right)
e^{-ik_Fx\cos\theta}
\end{align}

\begin{align} \nonumber
\psi_4(L+d< x<L+2d) = \frac{1}{\sqrt{2(E+U_2)(E+U_2-M_{z2})}} \nonumber \\
\cdot \Bigg[ f\left( \begin{array}{c}
\hbar v_F (\tilde{k}_{y2}+i\tilde{k}_{x2}) \nonumber \\
E+U_2-M_{z2}
\end{array} \right)
e^{i(\tilde{k}_{x2} + m_{y2}/\hbar v_F)x}\nonumber \\
+g\left( \begin{array}{c}
\hbar v_F (\tilde{k}_{y2}-i\tilde{k}_{x2}) \nonumber \\
E+U_2-M_{z2}
\end{array} \right)
e^{i(-\tilde{k}_{x2} + m_{y2}/\hbar v_F)x}\Bigg]
\end{align}

\begin{align} \nonumber
\psi_5(L+2d<x) = \frac{t}{\sqrt{2}}
\left( \begin{array}{c}
ie^{-i\theta} \nonumber \\
1
\end{array} \right)
e^{ik_Fx\cos\theta}
\end{align}
where $r$ is the reflection coefficient in region I, $t$ is the transmission
coefficient in region V, $a$, $b$, $c$, $d$, $f$, and $g$ are the corresponding
coefficients in regions II, III, and IV. Other quantities are
\begin{eqnarray}
\nonumber
E & = & \hbar v_F k_F, \\ \nonumber
k_y & = & k_F \sin \theta, \\ \nonumber
\hbar v_F \tilde{k}_{x1} &=& \sqrt{E^2 -m_{z1}^2 - (\hbar v_F \tilde{k}_y)^2},
\\ \nonumber
\hbar v_F \tilde{k}_{x2} &=& \sqrt{E^2 -m_{z2}^2 - (\hbar v_F \tilde{k}_y)^2},
\\ \nonumber
\hbar v_F \tilde{k}_{y1} & = & \hbar v_F k_y + m_{x1}, \\ \nonumber
\hbar v_F \tilde{k}_{y2} & = & \hbar v_F k_y + m_{x2},
\end{eqnarray}
$U_1$ and $U_2$ are the biases applied on the two FMIs, respectively, and $\boldsymbol{m}=\xi\boldsymbol{M}$.
Matching the wavefunctions at the boundaries, we get all the coefficients
and hence the wavefunction in the whole domain, as follows.
\begin{align} \nonumber
t = \frac{Z_5 Z_{10} - Z_6 Z_9}{(Z_4 Z_{10} - Z_6 Z_8)b_0 - (Z_6 Z_7 - Z_3 Z_10)a_0 },
\end{align}

\begin{align} \nonumber
r = \frac{(Z_5 Z_7 - Z_3 Z_9)a_0 - (Z_4 Z_9 - Z_5 Z_8)b_0}{(Z_4 Z_{10} - Z_6 Z_8)b_0 - (Z_6 Z_7 - Z_3 Z_{10})a_0},
\end{align}

\begin{align} \nonumber
a = \frac{(Y_6 Z_1 - Y_7 Y_{10})C_0 + (Y_6 Z_2 - Y_8 Y_{10})d_0}{Y_6 Y_9 - Y_5 Y_{10}} t=a_0 t,
\end{align}

\begin{align} \nonumber
b = \frac{(Y_6 Y_9 - Y_5 Z_1)C_0 + (Y_8 Y_9 - Y_5 Z_2)d_0}{Y_6 Y_9 - Y_5 Y_{10}} t=b_0 t,
\end{align}

\begin{align} \nonumber
c = \frac{(X_{10} Y_1 - X_7 Y_4)f_0 + (X_{10} Y_2 - X_8 Y_4)g_0}{X_{10}Y_3 - X_9Y_4}t = c_0 t,
\end{align}

\begin{align} \nonumber
d = \frac{(X_7 Y_3 - X_9 Y_1)f_0 + (X_8 Y_3 - X_9 Y_2)g_0}{X_{10}Y_3 - X_9Y_4}t = d_0 t,
\end{align}

\begin{align} \nonumber
f = \frac{X_2 X_6 - X_3 X_5}{X_2 X_4 - X_1 X_5}t = f_0 t,
\end{align}

\begin{align} \nonumber
g = \frac{X_3 X_4 - X_1 X_6}{X_2 X_4 - X_1 X_5}t = g_0 t,
\end{align}
where th variables $X_1$-$X_{10}$, $Y_1$-$Y_{10}$, and $Z_1$-$Z_{10}$ are
\begin{align} \nonumber
X_1 \equiv \hbar v_F (\tilde{k}_{y2} + i\tilde{k}_{x2})e^{i(\tilde{k}_{x2} + m_{y2}/\hbar v_F)(L+2d)},
\end{align}
\begin{align} \nonumber
X_2 \equiv \hbar v_F (\tilde{k}_{y2} - i\tilde{k}_{x2})e^{i(-\tilde{k}_{x2} + m_{y2}/\hbar v_F)(L+2d)},
\end{align}
\begin{align} \nonumber
X_3 \equiv ie^{-i\theta} e^{ik_F (L + 2d)\cos \theta} \sqrt{(E+U_2)(E + U_2 - m_{z2})},
\end{align}
\begin{align} \nonumber
X_4 \equiv (E+U2 - m_{z2})e^{i(\tilde{k}_{x2} + m_{y2}/\hbar v_F)(L+2d)},
\end{align}
\begin{align} \nonumber
X_5 \equiv (E+U2 - m_{z2})e^{-i(\tilde{k}_{x2} + m_{y2}/\hbar v_F)(L+2d)},
\end{align}
\begin{align} \nonumber
X_6 \equiv e^{ik_F (L + 2d)\cos \theta} \sqrt{(E+U_2)(E + U_2 - m_{z2})},
\end{align}
\begin{align} \nonumber
X_7 = e^{-ik_F (L + d)\cos \theta} \sqrt{(E+U_2)(E + U_2 - m_{z2})},
\end{align}
\begin{align} \nonumber
X_8 \equiv \hbar v_F (\tilde{k}_{y2} - i\tilde{k}_{x2})e^{i(-\tilde{k}_{x2} + m_{y2}/\hbar v_F)(L+d)},
\end{align}
\begin{align} \nonumber
X_9 \equiv ie^{-i\theta} e^{ik_F (L + 2d)\cos \theta} \sqrt{(E+U_2)(E + U_2 - m_{z2})},
\end{align}
\begin{align} \nonumber
X_{10} \equiv -ie^{-i\theta} e^{ik_F (L + 2d)\cos \theta} \sqrt{(E+U_2)(E + U_2 - m_{z2})},
\end{align}
\begin{align} \nonumber
X_{10} \equiv -ie^{-i\theta} e^{ik_F (L + 2d)\cos \theta} \sqrt{(E+U_2)(E + U_2 - m_{z2})},
\end{align}
\begin{align} \nonumber
Y_1 \equiv (E+U2 - m_{z2})e^{i(\tilde{k}_{x2} + m_{y2}/\hbar v_F)(L+d)},
\end{align}
\begin{align} \nonumber
Y_2 \equiv (E+U2 - m_{z2})e^{i(-\tilde{k}_{x2} + m_{y2}/\hbar v_F)(L+d)},
\end{align}
\begin{align} \nonumber
Y_3 \equiv e^{ik_F (L + d)\cos \theta} \sqrt{(E+U_2)(E + U_2 - m_{z2})},
\end{align}
\begin{align} \nonumber
Y_4 \equiv e^{-ik_F (L + d)\cos \theta} \sqrt{(E+U_2)(E + U_2 - m_{z2})},
\end{align}
\begin{align} \nonumber
Y_5 \equiv \hbar v_F (\tilde{k}_{y1} + i\tilde{k}_{x1})e^{i(\tilde{k}_{x1} + m_{y1}/\hbar v_F)d},
\end{align}
\begin{align} \nonumber
Y_6 \equiv \hbar v_F (\tilde{k}_{y1} - i\tilde{k}_{x1})e^{i(-\tilde{k}_{x1} + m_{y1}/\hbar v_F)d},
\end{align}
\begin{align} \nonumber
Y_7 \equiv ie^{-i\theta} e^{ik_F d\cos \theta} \sqrt{(E+U_1)(E + U_1 - m_{z1})},
\end{align}
\begin{align} \nonumber
Y_8 \equiv -ie^{-i\theta}e^{ik_F d\cos \theta} \sqrt{(E+U_1)(E + U_1 - m_{z1})},
\end{align}
\begin{align} \nonumber
Y_9 \equiv (E+U1 - m_{z1})e^{i(\tilde{k}_{x1} + m_{y1}/\hbar v_F)d},
\end{align}
\begin{align} \nonumber
Y_10 \equiv (E+U1 - m_{z1})e^{i(-\tilde{k}_{x1} + m_{y1}/\hbar v_F)d},
\end{align}
\begin{align} \nonumber
Z_1 \equiv e^{ik_F d\cos \theta} \sqrt{(E+U_1)(E + U_1 - m_{z1})},
\end{align}
\begin{align} \nonumber
Z_2 \equiv e^{-ik_F d\cos \theta} \sqrt{(E+U_1)(E + U_1 - m_{z1})},
\end{align}
\begin{align} \nonumber
Z_3 \equiv \hbar v_F (\tilde{k}_{y1} + i\tilde{k}_{x1}),
\end{align}
\begin{align} \nonumber
Z_4 \equiv \hbar v_F (\tilde{k}_{y1} - i\tilde{k}_{x1}),
\end{align}
\begin{align} \nonumber
Z_5 \equiv -Z_6 = ie^{-i\theta} \sqrt{(E+U_1)(E + U_1 - m_{z1})},
\end{align}
\begin{align} \nonumber
Z_7 \equiv Z_8 = E+U1 - m_{z1},
\end{align}
\begin{align} \nonumber
Z_9 \equiv Z_{10} = \sqrt{(E+U_1)(E + U_1 - m_{z1})},
\end{align}


%
\end{document}